\documentclass{aa}
\usepackage{graphicx}
\usepackage{natbib} 
\bibpunct{(}{)}{;}{a}{}{,}
\usepackage{txfonts}
\usepackage{siunitx}
\usepackage{enumitem}
\usepackage[]{hyperref}
\graphicspath{{graphics/}}
 
 \def\Halpha{$\mathrm{H\alpha}$}  \def\Hgamma{$\mathrm{H\gamma}$} \def\Hepsilon{$\mathrm{H\epsilon}$}
  \def\HII{\ion{H}{ii}} \def\HeI{\ion{He}{i}}
 \def\HeII{\ion{He}{ii}}  
   \def\CIV{\ion{C}{iv}}
    
 \def\NIII{\ion{N}{iii}} \def\NIV{\ion{N}{iv}} \def\NV{\ion{N}{v}}
   
  \def\OIV{\ion{O}{iv}} \def\OV{\ion{O}{v}}
 \def\OVI{\ion{O}{vi}}  
  \def\SiIV{\ion{Si}{iv}}

\newcommand{\msunpyr}{\ifmmode{\,M_{\odot}\,\mbox{yr}^{-1}} \else{ M$_{\odot}$/yr}\fi}

\newcommand{\kms}{\ifmmode{\,\mbox{km}\,\mbox{s}^{-1}}\else{km\,s^{-1}}\fi}
\newcommand{\kpc}{\ifmmode {\,\mbox{kpc}} \else{kpc}\fi}
\newcommand{\msun}{\ifmmode M_{\odot} \else M$_{\odot}$\fi}
\newcommand{\rsun}{\ifmmode R_{\odot} \else R$_{\odot}$\fi}
\newcommand{\lsun}{\ifmmode L_{\odot} \else L$_{\odot}$\fi}
\newcommand{\zsun}{\ifmmode Z_{\odot} \else $Z_{\odot}$\fi}
\newcommand{\xsun}{\ifmmode X_{\odot} \else $X_{\odot}$\fi}
\newcommand{\velo}{\ifmmode\varv\else$\varv$\fi}
\newcommand{\vinf}{\ifmmode\velo_\infty\else$\velo_\infty$\fi}
\newcommand{\rgal}{\ifmmode \,R_{\mathrm{gal}} \else R$_{\mathrm{gal}}$\fi}

\begin{document} 
 
\title{The  earliest O-type eclipsing binary in the Small Magellanic Cloud, AzV\,476: A comprehensive analysis reveals surprisingly low stellar masses}

\subtitle{}
\titlerunning{The  earliest O-type eclipsing binary in the Small Magellanic Cloud, AzV\,476.}

\author{D. Pauli$^1$ \and L.\,M. Oskinova$^1$ \and W.-R. Hamann$^1$ \and  V. Ramachandran$^{1,3}$ \and H. Todt$^1$ \and A. A. C. Sander$^{2,3}$ \and T. Shenar$^4$ \and M. Rickard$^{1,5}$ \and J. Maíz Apellániz$^6$ \and R. Prinja$^5$ 
}

\authorrunning{D. Pauli et al.}
 
\institute{Institut f{\"u}r Physik und Astronomie, Universit{\"a}t Potsdam, Karl-Liebknecht-Str. 24/25, 14476 Potsdam, Germany\label{inst1}
\and Armagh Observatory and Planetarium, College Hill, Armagh BT61 9DG, Northern Ireland, UK\label{inst2}
\and Astronomisches Rechen-Institut, Zentrum für Astronomie der Universität Heidelberg, Mönchhofstr. 12-14, 69120 Heidelberg, Germany
\and Institute  of  Astronomy,  KU  Leuven,  Celestijnenlaan  200D,  3001 Leuven, Belgium\label{inst3}
\and Department of Physics and Astronomy, University College London, Gower Street, London WC1E 6BT, UK\label{inst4}
\and Centro de Astrobiología. CSIC-INTA. Campus ESAC. Camino bajo del castillo s/n. E-28 692 Villanueva de la Cañada, Madrid, Spain.
}
 
\date{Received ; Accepted}

\abstract{Massive stars at low metallicity are among the main feedback
agents in the early Universe and in present-day star forming  galaxies. 
When in binaries, these stars are potential progenitors of gravitational-wave 
events. Knowledge of stellar masses is a prerequisite to understanding evolution and feedback of
low-metallicity massive stars.} 
{Using abundant spectroscopic and photometric measurements of an outstandingly bright 
eclipsing binary, we compare its dynamic, spectroscopic, and evolutionary mass estimates and develop  
a binary evolution scenario.
}
{We comprehensively studied the eclipsing binary system, AzV\,476, in the Small Magellanic Cloud (SMC). The
light curve and radial velocities were analyzed to obtain the orbital parameters. The photometric and spectroscopic data in the UV and optical were analyzed using the Potsdam Wolf-Rayet (PoWR) model atmospheres. The obtained results are interpreted using detailed binary-evolution tracks including mass transfer.}
{AzV\,476 consists of an O4\,IV-III((f))p primary and an O9.5:\,Vn secondary.
Both components have similar current masses (${20\,\msun}$ and ${18\,\msun}$) obtained consistently from both the orbital and spectroscopic analysis. The effective temperatures are $\SI{42}{kK}$ and $\SI{32}{kK}$, respectively. 
The  wind mass-loss rate of ${\log(\dot{M}/(\msunpyr))=-6.2}$ of the primary is a factor of ten higher than a recent empirical prescription for single O stars in the SMC.
Only close-binary evolution with mass transfer can  reproduce
the current stellar and orbital parameters, including orbital separation, eccentricity, and the rapid rotation of the secondary. The binary evolutionary model reveals that the primary has lost about half of its initial mass and is already core helium burning.}
{Our comprehensive analysis of AzV\,476 yields a consistent set of parameters and suggests previous case B mass transfer. The derived stellar masses agree within their uncertainties. The moderate masses of AzV\,476 underline the scarcity of bright massive stars in the SMC. The core helium burning nature of the primary indicates that stripped stars might be hidden among OB-type populations.}

\keywords{binaries: eclipsing - binaries: close - 
binaries: spectroscopy - stars: early-type - stars: fundamental parameters - 
stars: individual: AzV\,476}
\maketitle

\section{Introduction} 
\label{sec:intro} 

    The most important parameter defining the evolution of a star is its mass. An often reported problem in stellar astrophysics is the ``mass discrepancy'' problem, which refers to the inconsistent masses derived from spectroscopy, evolutionary tracks, and, for binaries, from orbital motions \citep{her1:92,wei1:10,mar1:15}. The analysis of O-type stars located in the Galaxy and the Large Magellanic Cloud (LMC) performed by \citet{wei1:10} reveal good agreement between the aforementioned mass estimates, but the study of  \citet{mar1:18} suggests a mass discrepancy. \citet{mah1:20} investigate a  sample of O-type binaries in the LMC and find good agreement between spectroscopic and dynamic mass estimates while the evolutionary masses are at odds. These latter authors suggest previous binary interactions as a possible solution for this discrepancy.
    
    However, the mass discrepancy problem has not yet been studied at metallicities lower than ${\lesssim1/2\,\zsun}$. Sufficiently low metallicity is offered by the nearby Small Magellanic Cloud (SMC) galaxy 
    ($Z_{\rm SMC}\approx1/7\,\zsun$;\citet[][]{hun1:07,tru1:07}). Precise stellar masses in the SMC allow us to tackle questions about stellar evolution and feedback at low metallicity. 
   
    Stars with spectral types around O2-4 are expected to be very massive with $M_\ast \gtrsim50\,\msun$ \citep{mar1:21}.
    However, the true masses of the early-type O stars are only poorly known, and therefore spectral types and masses might be falsely mapped. So far, only a couple of the SMC eclipsing binaries with early spectral types have been studied. \citet{mor1:03} investigated the O6V+O4/5III(f) system \mbox{Hodge~53-47} (alias MOA\,J010321.3-720538) and found dynamic masses of ${\approx26\,\msun}$ and ${\approx16\,\msun}$ for the primary and secondary, respectively. The eclipsing binary, \mbox{OGLE~SMC-SC10~108086}, studied by \citet{abd1:21} contains even less massive stars with ${\approx17\,\msun}$ and ${\approx14\,\msun}$. Only one eclipsing multiple stellar system in the SMC, HD\,5980, appears to contain stars with masses $\gtrsim30\,\msun$. \citet{koe2:14} and \citet{hil1:19} studied this system in detail and found an inner eclipsing binary consisting of an LBV and a Wolf-Rayet (WR) star, and a third O-type supergiant star with a potential fourth companion. The orbital masses of the LBV and the WR star are ${\approx61\,\msun}$ and ${\approx66\,\msun}$, respectively. \citet{koe2:14} suggest that there was little or no mass transfer, and that the WR star has formed via quasi-chemically homogeneous evolution.
    These examples highlight the complexity of massive star evolution and the need for advanced studies on the mass discrepancy problem in the high-mass regime.
    
    \begin{table*}[tbp]
        \centering
        \caption{List of all spectra of AzV\,476 used in this work and their associated orbital phase. RVs and the wavelengths ranges and lines used for their measurements are also listed. The listed RVs are already corrected for the barycentric motion and the velocity of the NGC 456 complex  with $\varv_\mathrm{SMC}= \SI{152}{km\,s^{-1}}$.}
        
        \setlength{\tabcolsep}{4pt}
        \begin{tabular}{cc|ccc|cc}\hline \hline \rule{0cm}{2.2ex}%
                spectral ID & Instrument & Wavelength                & MJD$^{(a)}$& Phase $\phi$ $^{(b)}$ & RV$_1$        & RV$_2$   \\
                       &            & [\AA]                            & [d]      &         &[$\si{km\,s^{-1}}$]&[$\si{km\,s^{-1}}$] \\
                \hline \rule{0cm}{2.2ex}%
                1      & FUSE       & $\,\,\,\SIrange{950}{1150}{\AA}$ & 52478.3849 &    $\,\,\,\,0.0192$  & ---        & ---             \\
                2      & HST/COS    & $\SIrange{1178}{1777}{\AA}$      & 56185.2387 &    $-0.2305$  & $\,\,\,\,\,194\pm\,\,5\,^{(c)}$  & ---          \\
                3      & HST/STIS   & $\SIrange{1140}{1735}{\AA}$      & 59022.5673 &    $-0.3124$  & $\,\,\,\,\,\,\,\,91\pm\,\,5\,^{(c)}$       & $\,\,\,\,\,\,\,\,\,-86\pm12\,^{(c)}$ \\
                4      & HST/STIS   & $\SIrange{1574}{2673}{\AA}$      & 57321.6698 & $\,\,\,\,0.0968$ & $\,\,\,\,-55\pm\,\,5\,^{(d)}$  & $\,\,\,\,\,\,\,\,\,\,\,\,\,64\pm45\,^{(d)}$ \\
                5      & X-SHOOTER   & $\SIrange{3000}{10000}{\AA}$     & 59163.2501 & $-0.2928$ &  $112\pm14$  &  $-110\pm10$  \\
                6      & UVES       & $\SIrange{3000}{11000}{\AA}$     & 57703.2868 & $-0.1610$ &  $223\pm\,\,9\,$  & $-171\pm15$  \\
                7      & UVES       & $\SIrange{3000}{11000}{\AA}$     & 57748.0479 & $-0.3823$ &  $\,\,\,60\pm16$   &  $\,\,\,-42\pm17$    \\
                8      & UVES       & $\SIrange{3000}{11000}{\AA}$     & 57749.0558 & $-0.2747$ &  $126\pm\,\,8\,$  &  $-104\pm29$   \\
                9      & UVES       & $\SIrange{3000}{11000}{\AA}$     & 57749.0923 & $-0.2708$ &  $141\pm16$  &  $\,\,\,-95\pm37$ \\
                10     & UVES       & $\SIrange{3000}{11000}{\AA}$     & 57750.1132 & $-0.1618$ &  $201\pm13$  &  $-184\pm34$  \\
                11     & UVES       & $\SIrange{3000}{11000}{\AA}$     & 57750.1492 & $-0.1580$ &  $207\pm\,\,9\,$  &  $-186\pm43$  \\
                12     & UVES       & $\SIrange{3000}{11000}{\AA}$     & 57751.0782 & $-0.0588$ &  $148\pm12$  &  $\,\,\,-71\pm37$  \\
                \hline
        \end{tabular}
        \rule{0cm}{4ex}%
        \begin{minipage}{0.85\linewidth}
            \ignorespaces
            $^{(a)}$ Mid-exposure in HJD - 2400000.5 $^{(b)}$ Calculated with Eq.~(\ref{eq:phase}). $^{(c)}$ Obtained from a fit over the range $\SIrange{1360}{1405}{\AA}$. $^{(d)}$ Obtained from a fit over the range $\SIrange{2100}{2230}{\AA}$.%
        \end{minipage}
        \label{tab:spectroscopy_and_RV}
    \end{table*}

    For single stars, mass estimates rely on spectroscopic diagnostics or comparison with evolutionary tracks. 
    Currently, standard stellar evolutionary models predict that the most massive stars, either single or in binary systems, start their lives on the main sequence (MS) as early-type O stars.  Single stars expand and evolve away from the MS. During this phase, stars undergo strong mass loss where they might lose their entire hydrogen-rich envelope, revealing the helium core and becoming WR stars. In the case of close binaries, evolutionary models predict that the expanding star is likely to interact with its companion. 
    In the majority of cases, the expanding star will transfer its envelope to the companion, and become a binary-stripped helium star, possibly also with a WR-type spectrum \citep{Dionne2006,She1:20,Goe1:20}.

    The SMC hosts a handful of WR stars that have relatively high masses ranging from $10\,\msun$ to $60\,\msun$ \citep{she1:16}. These stars are so massive that their hydrogen-burning progenitors must have been early O-type (or WNL/Of) stars. 
    However, in their recent study of the SMC OB-type population, \citet{ram1:19} reveal a strong deficiency of  massive stars ($>30\,\msun$) close to the MS in the upper part of the empiric  Hertzsprung–Russell diagram (HRD) \citep[see][for studies of Galactic and the LMC O star populations]{hol1:20, ram1:18}. O-type stars are hot and luminous and are therefore easily detectable. Hence, a paucity of the earliest O-type stars in the SMC cannot be explained either by selection effects or stellar evolution scenarios \citep{sch1:21}. Thus, the deficiency of most massive O stars  strongly questions the formation process of the WR stars in the SMC, as well as our basic understanding of stellar evolution in low-metallicity environments. 
    In this context it is crucial to quantify the budget of massive stars ($\gtrsim30\,\msun$) in the SMC.
    
    Poorly constrained processes that affect the lives of massive stars include
    radiatively driven winds, stellar envelope inflation, core overshooting, and rotationally induced internal mixing. In particular their dependence on mass and metallicity are not yet fully understood. These effects can drastically alter the evolution of a star. 
    Studies of objects in low-metallicity environments are needed to 
    constrain these metallicity-dependent effects and thus allow the establishment of reliable stellar evolutionary tracks.
    
    To address these outstanding questions, we selected one of the earliest subtype O stars in the SMC that is an eclipsing binary, allowing us to estimate its mass by various methods. 
    The subject of this paper, AzV\,476, is located in the cluster NGC\,456 in the SMC Wing. The cluster contains an active star-forming region and hosts young stellar objects \citep{mur1:17}.  AzV\,476 is embedded in a \HII{} region. The system was identified as an eclipsing binary by the Optical Gravitational Lensing Experiment (OGLE), which monitors stellar variability in the SMC and LMC \citep{paw1:16}.
    The primary star was classified  by \citet{mas1:05} as an O2-3\,V star based on its optical spectral appearance and is therefore one of the earliest type stars in the entire SMC. Here we present the first consistent analysis of AzV\,476, based on photometric and spectroscopic data, and in particular accounting for its binary nature. Newly obtained high-resolution UV and optical data give us the opportunity to perform such an analysis, yielding estimates on masses and stellar and wind parameters.  
    
    This paper is structured as follows. In Sect.~\ref{sec:Observations} we describe the observations and the known stellar and orbital parameters used in the analysis below. The results of the orbital and spectral analysis as well as those from evolutionary modeling are presented in Sects.~\ref{sec:orbit_analyis},~\ref{sec:spectral_analyis},~and~\ref{sec:evol_analysis}, respectively. Their implications, similarities, and disagreements on the stellar masses and other stellar parameters are discussed in Sect.~\ref{sec:disc}, and conclusions are given in Sect.~\ref{sec:conclusions}.

\section{Observations}
\label{sec:Observations}

\subsection{Spectroscopy}

    Over the last decade, a handful of AzV\,476 spectra covering the UV, optical (VIS), and infrared (IR) have been obtained, yielding multi-epoch data well suited for measuring radial velocities (RVs) and spectroscopic analysis. We used a spectroscopic dataset consisting of 12 spectra and  Table~\ref{tab:spectroscopy_and_RV} gives a brief overview of these, their covered wavelength ranges, their observed date, and their associated orbital phase.
    In the remainder of this paper, we refer to the individual observed spectra by their ID in Table~\ref{tab:spectroscopy_and_RV}.
    
    The wavelength regime  $\SIrange{950}{1150}{\AA}$ is covered by an archival  FUSE \citep{oeg1:00} observations\footnote{the shorter wavelengths in the FUSE range are largely contaminated by interstellar features that yield no information about the stellar parameters}.
    The FUSE spectrum (ID 1) was taken with a total exposure time of ${\SI{21567}{s}}$ and a resolving power of ${R\approx20\,000}$. Incidentally, the FUSE observation was taken close to the primary eclipse. This FUSE spectrum has a known but unsolved calibration issue. Therefore, we only used the LiF1A, LiF2A, and SiC2A channels, which appear to be the least affected by this problem. The spectrum is rectified by division through the combined continuum flux of our models.
    
    AzV\,476 is part of the ULLYSES program\footnote{https://ullyses.stsci.edu/}. It was observed with the HST/COS \citep{hir1:21} using the G130M (${\SIrange{1178}{1472}{\AA}}$) and G160M (${\SIrange{1383}{1777}{\AA}}$) medium-resolution gratings (ID 2), ${R\approx19\,000}$. The two spectra were taken sequentially with exposure times of ${\SI{330}{s}}$ and ${\SI{1100}{s}}$, respectively.
In addition, the star was re-observed in the UV with the HST/STIS spectrograph \citep{bra1:21} using the E140M echelle graiting  (ID 3) as part of the HST program 15837 (PI Oskinova). The spectrograph covers a wavelength regime of ${\SIrange{1140}{1735}{\AA}}$. The exposure time was $\SI{2707}{s}$ and the final resolving power is ${R\approx45\,800}$.
    
    There is another spectrum in the ULLYSES program (ID 4) taken with the HST/STIS spectrograph using the E230M echelle gratings covering a wavelength range ${\SIrange{1574}{2673}{\AA}}$. The exposure time was ${\SI{2820}{s}}$ and the resolving power is ${R=30\,000}$.
    
    For the optical and near-IR range, we use the publicly available spectra taken with the X-SHOOTER spectrograph \citep{ver1:11} mounted on the ESO Very Large Telescope (VLT). The spectrum (ID 5) was taken as part of the ESO 106.211Z program, which is part of the XSHOOTU program (their Paper I; Vink and the XShootU Collaboration, in prep.). The X-SHOOTER spectrograph consists of three different spectroscopic arms, which are optimized for the wavelength ranges in the UBV (${\SIrange{3000}{5550}{\AA}}$), VIS (${\SIrange{5300}{10000}{\AA}}$), and near-IR (${\SIrange{10000}{25000}{\AA}}$). The resolving powers are ${R\approx6\,600}$, ${R\approx11\,000,}$ and ${R\approx8\,000}$, respectively.
    The spectra were obtained with exposure times of ${\SI{750}{s}}$, ${\SI{820}{s,}}$ and ${\SI{300}{s}}$ for the UBV, VIS, and NIR arm, respectively. 
    
    The remaining seven spectra (ID $\numrange{6}{12}$) used for our analysis are taken with the UVES spectrograph \citep{dek1:00} mounted on the ESO VLT. Each spectrum was taken with the DIC1 setup covering the wavelength ranges of ${\SIrange{3000}{4000}{\AA}}$ and ${\SIrange{5000}{11000}{\AA}}$. The exposure times are about ${\SI{2895}{s}}$ for both spectrographs and have a resolving power of ${R\approx65\,000}$ and ${R\approx75\,000}$, respectively. The UVES spectra were rectified by hand.

\subsection{Photometry}

    \begin{table}[t]
        \centering
        \caption{UBVRIJHK photometry of AzV\,476.}
        \begin{tabular}{cc}\hline \hline \rule{0cm}{2.2ex}%
                Band & Apparent magnitude\\
                     & [mag]\\
                \hline \rule{0cm}{2.2ex}%
                U & $12.49\pm0.04$ \\
                B & $13.54\pm0.07$ \\
                V & $13.48\pm0.01$ \\
                V$_\mathrm{OGLE}$ & $13.49$ \\
                R & $13.72\pm0.06$ \\
                I & $13.54\pm0.20$ \\
                I$_\mathrm{OGLE}$ & $13.56$ \\
                J & $13.70\pm0.03$ \\
                H & $13.72\pm0.04$ \\
                K & $13.86\pm0.06$ \\
                G & $13.51\pm0.06$ \\
                G$_\mathrm{BP}$ & $13.46\pm0.01$ \\
                G$_\mathrm{RP}$ & $13.57\pm0.01$ \\
                \hline
        \end{tabular}
        \label{tab:photometry}
    \end{table}   
    
    The UBI photometry is adopted from the catalog of the SMC stellar population \citep{bon1:10}. 
    For the VR photometry, we use the  magnitudes from the fourth United States Naval Observatory (USNO) CCD Astrograph Catalog (UCAC4) \citep{zac1:12}. For completeness, we compare the values of the V- and I-band magnitudes to those listed in the IVth OGLE Collection of Variable Stars \citep{paw1:16} (V$_\mathrm{OGLE}$ and I$_\mathrm{OGLE}$). Unfortunately, the OGLE magnitudes are published without error margins. Nonetheless, we find that they are in agreement with the V- and I-band magnitude from \citet{zac1:12} and \citet{bon1:10}.  JHK photometry is from the 2MASS catalog \citep{cut1:03}. Additionally, we use the recent EDR3  Gaia photometry \citep{gai1:16,gai1:21}. A total list of the used magnitudes is shown in Table~\ref{tab:photometry}. \citet{mas1:05}  estimated the extinction towards AzV\,476 to be $E_\mathrm{B-V} = \SI{0.28}{mag}$ based on averaging the color excesses in $B-V$ and $U-B$ based on the spectral type. From our spectral energy distribution (SED) fit we find better agreement when using a lower extinction of $E_\mathrm{B-V} = \SI{0.26}{mag}$ (see Sect.~\ref{sec:spec_analysis}).
    The slight difference can be due to the use of different reddening laws.
    
    For the light curve modeling, we use the OGLE I-band photometry from the IVth OGLE Collection of Variable Stars \citep{paw1:16}. The photometric data are taken over the period from May 2010 to January 2014. We do not use the OGLE V-band photometry because it contains too few data points and therefore cannot be used to resolve the eclipses. 

    AzV\,476 was observed by the TESS space telescope in 2018 (sector 2) and 2020 (sectors 27 and 28). However, the relatively low spatial resolution of TESS ($\SI{21}{''\,px^{-1}}$) precludes accurate point source photometry in the crowded region around our target. Therefore, the TESS data are not used for light-curve modeling. However, we use the TESS data to improve the ephemeris (see Sect.~\ref{sec:light_curve}).
    
\subsection{Distance and location}
\label{sec:distance_and_location}
    Our target, AzV\,476 is part of the NGC 456 cluster which is located in the SMC Wing.
    The distance to the SMC Wing was estimated by \citet{cig1:09} ${d\approx \SI{55}{kpc}}$ corresponding to a distance modulus of $\mathrm{DM} = \SI{18.7}{mag}$. This is in agreement with \cite{nid1:13} who use red clump stars to study the structure of the SMC, including the SMC Wing, and find that it follows a bimodal distribution with a near component at a distance of ${d\approx \SI{55}{kpc}}$ and a far component at ${d\approx \SI{67}{kpc}}$. \citet{tat1:20} confirm this distance, but they speculate that young structures, such as the NGC 456 cluster, do not trace substructures that are associated with the intermediate-age populations and might be located in front of them (see Sect.~\ref{sec:mass_discrepancy}). In this work we adopt a distance of ${d=\SI{55}{kpc}}$.
    
    The observed spectra of AzV\,476 are corrected for barycentric motion, which was calculated with the tool described in \citet{wri1:14}. Additionally, we shifted the spectra by the RV of the NGC\,456 complex in the SMC $\varv_\mathrm{SMC}= \SI{152}{km\,s^{-1}}$
    determined by fitting  Gaussians to several interstellar medium (ISM) lines that we
    associate with the environment of AzV\,476. The RV is not uniform throughout the different regions in the SMC and our finding is in agreement with the results of \citet{pro1:10}.

\section{Analysis of the binary orbit}
\label{sec:orbit_analyis}
\subsection{Method: Eclipse light-curve and RV curve modeling}
\subsubsection{Light curve}
\label{sec:light_curve}
    Our target, AzV\,476, is listed in the IVth OGLE Collection of Variable Stars \citep{paw1:16} with an orbital period of ${P_\mathrm{OGLE}=\SI{9.3663198}{d}}$ and an epoch of the primary eclipse of ${T_{0\mathrm{,\,\,OGLE}} = 2457002.7608}$ in Heliocentric Julian Date (HJD). 
    Unfortunately, the orbital period and the epoch of the primary eclipse are given without error margins. The observations that constitute the OGLE I-band light curve were taken with a cadence of $\approx\SI{2}{d}$, and therefore the individual eclipses are not well resolved. 
    In contrast, the TESS light curve has a much finer time coverage of $\approx\SI{5}{h,}$ allowing better constraint of the orbital period and the epoch of the primary eclipse.
    
    The date at which an eclipse occurs can be expressed as
    \begin{equation}
        T(n) = P \cdot n +{T}_{0},
    \end{equation}
    where $T$ is the time of the primary eclipse at orbital cycle $n$. We fitted Gaussians to the primary eclipses in the TESS light curve; the corresponding orbital cycles and dates of all eclipses are listed in Table~\ref{tab:TESS}.
    \begin{table}[]
        \centering
        \caption{Dates of the primary eclipses in the TESS light curve.}
        \begin{tabular}{cc}\hline \hline \rule{0cm}{2.2ex}%
            orbital cycle $n$ & MJD$^{(a)}$\\
                 & [d]\\
            \hline \rule{0cm}{2.2ex}%
            145 &  $58360.4633\pm0.0027$\\
            146 &  $58369.8256\pm0.0030$\\
            147 &  $58379.1905\pm0.0029$\\
            218 &  $59044.2165\pm0.0016$\\
            219 &  $59053.5942\pm0.0016$\\
            220 &  $59062.9590\pm0.0014$\\
            221 &  $59072.8303\pm0.0014$\\
            222 &  $59081.6913\pm0.0015$\\
            \hline
        \end{tabular}
        \rule{0cm}{6ex}%
        \begin{minipage}{0.995\linewidth}
            \ignorespaces
            $^{(a)}$ ${\mathrm{MJD} = \mathrm{HJD} - 2400000.5}$. The TESS data are given in ${\mathrm{TBJD} = \mathrm{BJD} -  2457000.0}$ and that we converted the $\mathrm{BJD}$ to $\mathrm{HJD}$ to be comparable to the OGLE data.%
        \end{minipage}
        \label{tab:TESS}
    \end{table}
    Using this procedure, we obtain ${P=\SI{9.36665 \pm 0.00025}{d}}$ and  ${{T}_{0} = 2457002.7968 \pm 0.0052}$.
    
\begin{figure}[tbp]
        \centering
        \includegraphics[trim= 1.0cm 0.4cm 0.5cm 0.4cm ,clip ,width=0.50\textwidth]{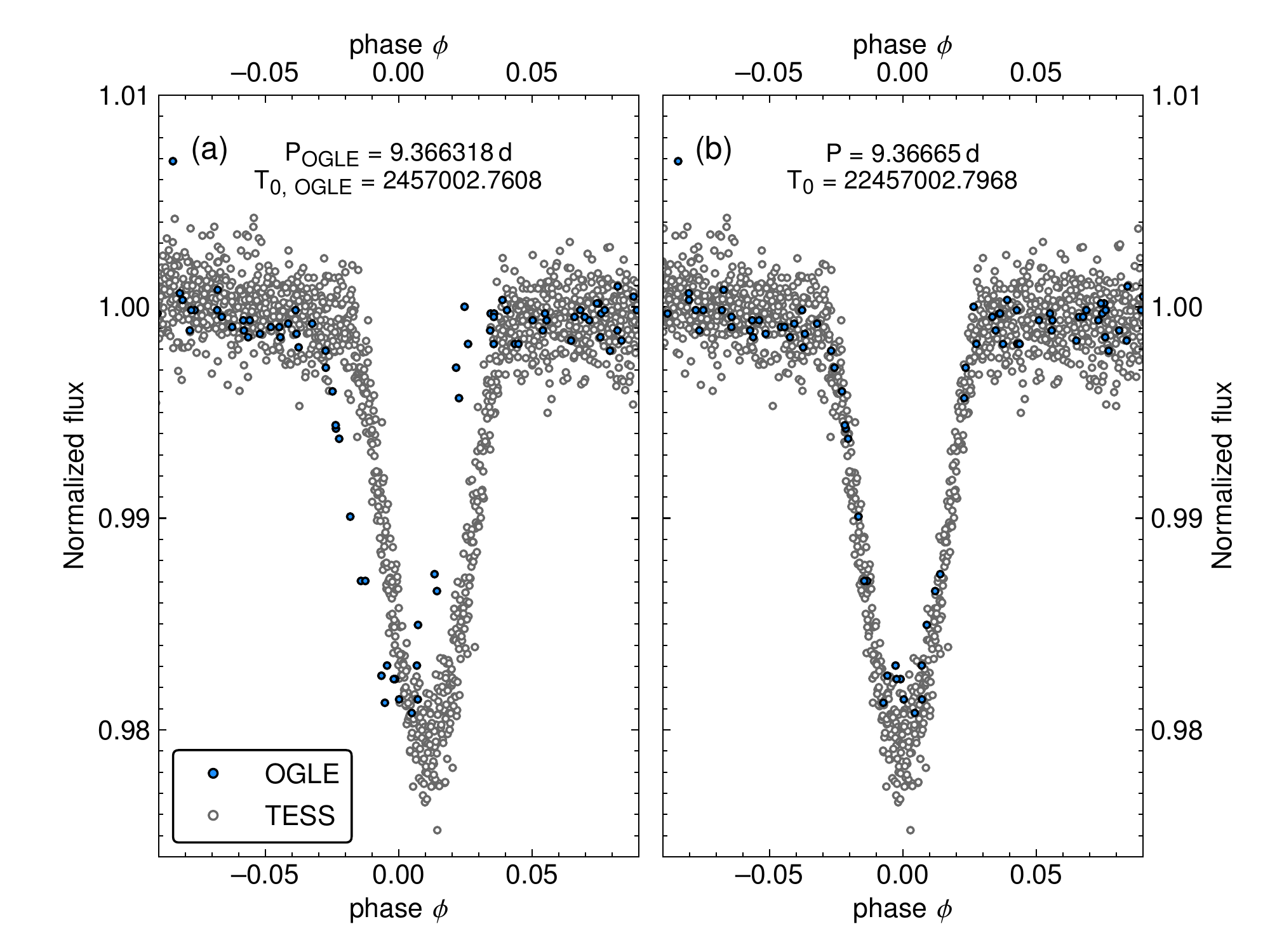}
        \caption{
        Phased OGLE I-band and TESS light curves around the primary eclipse. {\em Left panel:} Light curve phased according to the ephemeris from  the OGLE catalog.   {\em Right panel:} Light curve phased according to the ephemeris we derive in this work by fitting the primary's eclipses of the TESS light curve. }
        \label{fig:primary_eclipse}
    \end{figure}

    Figure~\ref{fig:primary_eclipse} shows a part of the phased light curve centered on the primary eclipse for the two different ephemerides, those published in the OGLE catalog and those we obtain    in this work. As can be seen,  our newly obtained ephemeris adequately describes the OGLE as well as the TESS light curves, i.e., the data set that covers $>\SI{10}{yr}$ of observations.

    In order to convert a date $t$ at which a spectrum was taken to a phase $\Phi,$ the following formula is used  
    \begin{equation}
        \phi(t) = 
        \begin{cases}
            \dfrac{t-{T}_0}{P}\mod{1}& \text{if } \phi<0.5\\
            \rule{0cm}{0.7cm}
            \dfrac{t-{T}_0}{P}\mod{1}-1  & \text{otherwise.}
        \end{cases}
        \label{eq:phase}
    \end{equation}
    
\begin{figure}[tbp]
        \centering
        \includegraphics[width=0.50\textwidth]{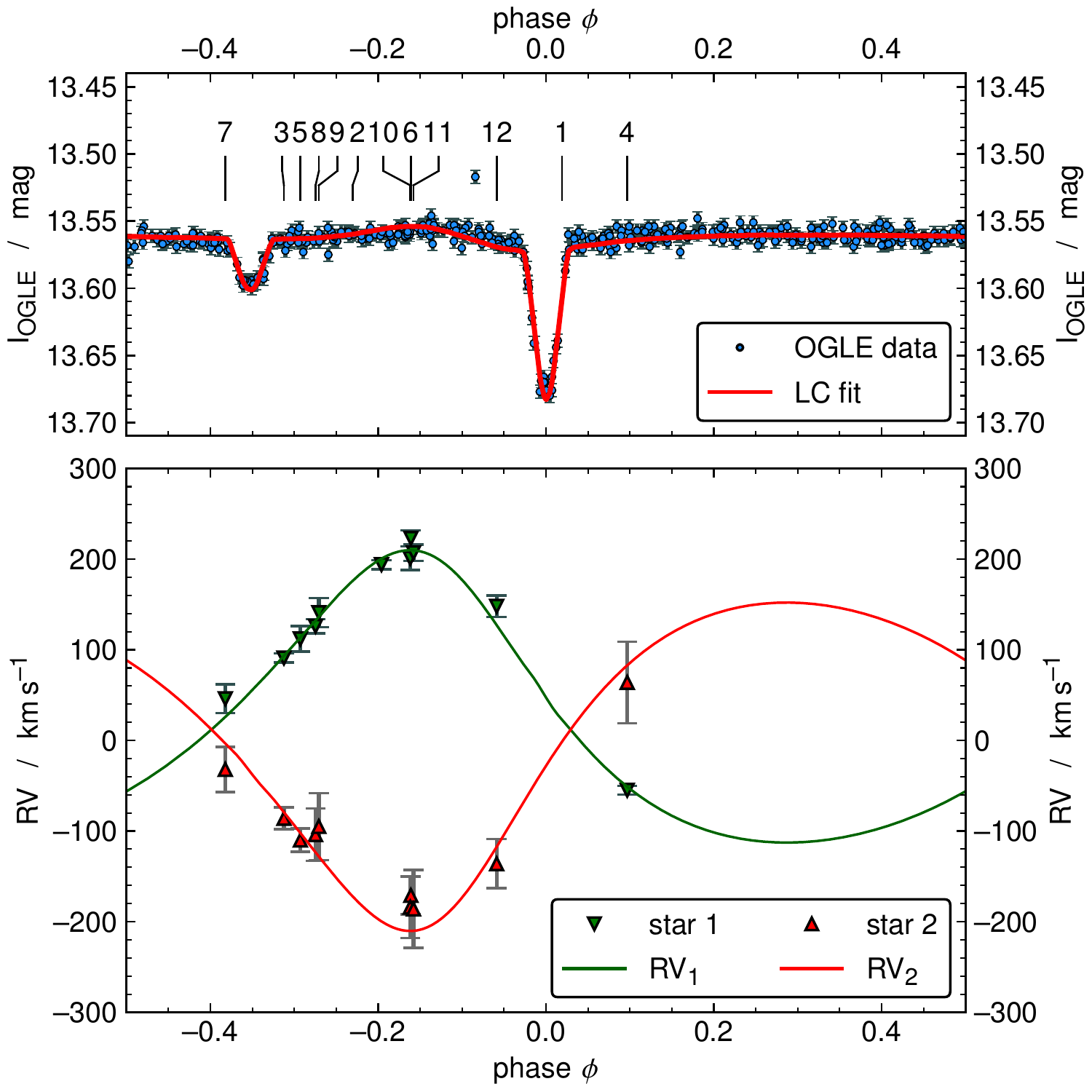}
        \caption{\textit{Upper panel:} Phased OGLE I-band light curve of AzV\,476 (blue dots) and the best fit obtained with the PHOEBE code (red line). Above the light curve, the spectral IDs of all used spectra (see Table~\ref{tab:spectroscopy_and_RV}) are indicated. \textit{Lower panel:}
        Observed (triangles) and fitted (solid lines) RV curves for the primary (green) and secondary (red). The fits of the light-curve fit and the RVs are consistently obtained by the PHOEBE code.}
        \label{fig:LC_and_RV_fit}
    \end{figure}

    With the improved ephemerides and their small error margins, the uncertainties on the phases are negligible and are therefore not taken into account here.
The OGLE I-band light curve is shown in the upper panel of Fig.~\ref{fig:LC_and_RV_fit}. The  conjunctions are at phases ${\phi=0.0}$ and ${\phi=-0.34,}$ implying that the orbit is eccentric.
    
\subsubsection{ RVs}

    To measure the RVs, we use a Markov chain Monte Carlo (MCMC) method combined with a least-square fitting method. In the MCMC method, the individual synthetic spectra of the primary and secondary (see Sect.~\ref{sec:spec_analysis}) are shifted by different RVs. The combined synthetic spectrum is then compared to the observation. Using a least-square likelihood function we estimate the quality of the used RVs. The MCMC method quickly explores a large parameter space of different RVs until it converges toward the true solution. In the vicinity of the true solution, the MCMC method calculates the probabilities of different combinations of the RVs. This yields the final probability distribution around the true solution. A more detailed explanation is given in Appendix~\ref{app:RV}.
    
    Because the final probability distribution obtained with the MCMC method is not necessarily a Gaussian, we quote the error as the 68\%\ confidence interval. The PHOEBE code cannot handle asymmetric errors, and therefore we only use the larger margin of the probability distribution as it is the safer choice. The uncertainties are included in the eclipsing binary modeling as described in Sect.\,\ref{sec:phoebe}.

    The primary dominates the emission and absorption lines. In order to avoid uncertainties due to the intrinsic variability of lines formed in the stellar wind, the RVs listed in Table~\ref{tab:spectroscopy_and_RV} are the averaged values of the RVs obtained from selected individual lines. A more detailed list of the RVs obtained from the individual lines in the different optical spectra is given in Tables~\ref{tab:RV1} and \ref{tab:RV2}.
    
\begin{figure}[ht]
        \centering
        \includegraphics[trim= 0.0cm 0cm 0.cm -0.4cm ,clip ,height=0.5\textwidth,angle = -90]{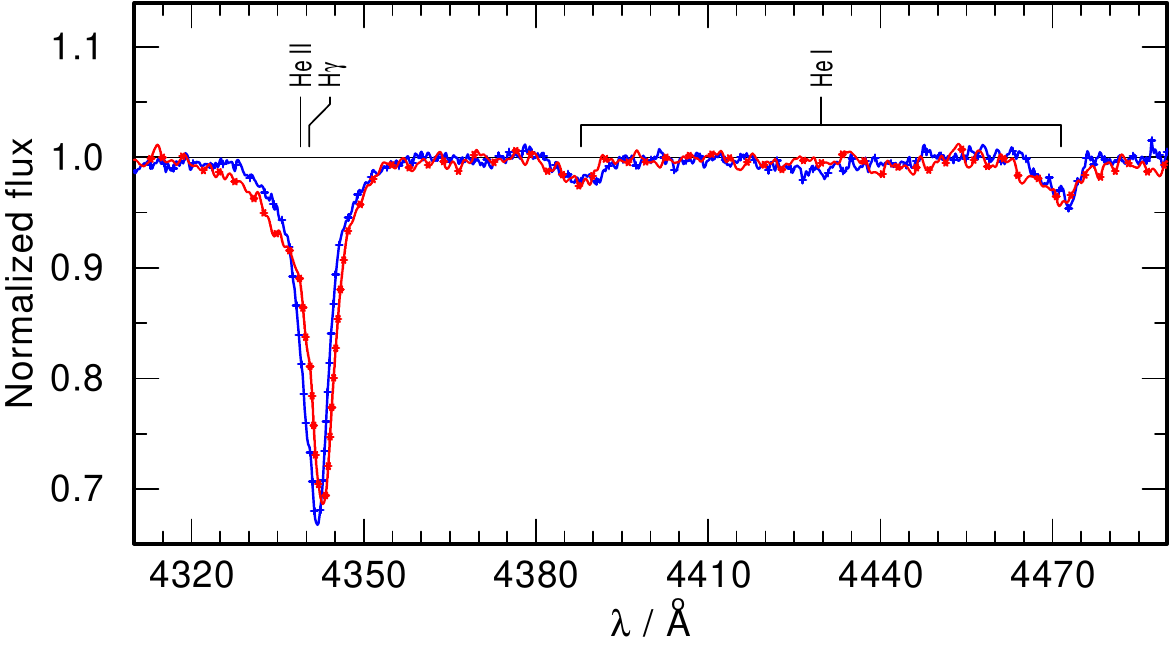}
        \caption{
        X-SHOOTER spectrum (ID 5 in Table~\ref{tab:spectroscopy_and_RV}) displayed in blue and one of the UVES spectra (ID 10 in Table~\ref{tab:spectroscopy_and_RV}) in red. The spectra are convolved with a Gaussian with an ${\mathrm{FWHM}=\SI{0.4}{\AA}}$ to reduce the noise and to make the RV shifts visible.
        The region containing \Hgamma{}, \HeI{} $\lambda4387,$ and \HeI{} $\lambda4471$ lines is shown. In the spectrum shown by the red line,  the primary's spectrum is redshifted (see \Hgamma{}), while  the secondary's spectrum (broadened \HeI{} lines) is blueshifted. We note that the primary also partially contributes to the \HeI{} lines.}
        \label{fig:rv_shifts}
    \end{figure}

    All the absorption lines that are associated with the secondary show a contribution from the primary. Furthermore, the depths of these absorption lines are at the level of the noise, which introduces additional uncertainties; these are reflected in the larger error margins.
    Two selected optical spectra obtained at different phases are depicted in 
    Fig.\,\ref{fig:rv_shifts} to demonstrate how the spectral lines associated with the different binary components shift.
    
\subsubsection{Modeling with PHOEBE}
    
\label{sec:phoebe}
    The Physics of Eclipsing Binaries (PHOEBE) v.2.3 modeling software \citep{prs1:05,prs1:16,hor1:18,jon1:20,con1:20} is employed to derive dynamic masses as well as to obtain measures of the stellar parameters independently from the spectroscopic model. The simultaneous fitting of the RV and light curve is done with the \verb|emcee| sampler \citep{for1:13}.
    
    To reduce the parameter space, we fix the orbital period and the epoch of the  primary eclipse to the values we derive from the TESS light curve; see Sect.~\ref{sec:light_curve}. As we are using only one passband, the temperatures of the components cannot be obtained reliably. Therefore, the temperatures of the primary and secondary are fixed to the results from the spectral analysis (see Sect.~\ref{sec:spec_analysis}), ${T_\mathrm{eff,\,1} = \SI{42}{kK}}$ and ${T_\mathrm{eff,\,2} = \SI{32}{kK}}$, respectively.
    
    The PHOEBE code assumes synchronous stellar rotation. The actual fast rotation of the secondary with $\varv \sin i = \SI{425}{km\,s^{-1}}$
    is therefore not consistently taken into account for modeling the secondary eclipse.  Gravitational darkening is modeled by a power law with a coefficient of  $\beta_\mathrm{grav}=1$, as recommended for radiative envelopes in the PHOEBE documentation\footnote{www.phoebe-project.org}. Given the period of $\sim\SI{10}{d}$, this renders gravitational darkening unimportant.

    The atmospheres of both components are approximated by a blackbody. We compared the blackbody flux to the SED of our atmospheric model and find that it is a valid approximation for the I-band flux. We calculated the emergent flux distribution using our spectral models (see Sect.~\ref{sec:spec_analysis}) and fitted different types of limb-darkening laws \citep[e.g.,][]{dia1:92}. We find that the limb-darkening law that
best describes the primary and secondary is a quadratic approximation in the form of
    \begin{equation}
        I(\mu)=I(1)\,[1-a_i\,(1-\mu)-b_i\,(1-\mu)^2],
    \end{equation}
    where $\mu=\cos \theta$ is the cosine of the directional angle $\theta$, and $a_i$ and $b_i$ are the limb-darkening coefficients of each stellar component $i$. For the primary, the best fit is achieved with coefficients $a_1=0.2032$ and $b_1=0.0275$, while for the secondary $a_2=0.1668$ and $b_2=0.0802$ are required.
    
    In a binary where both stellar components have large radii, on the order of $\gtrsim20\%$ of their separation, and rather similar temperatures, the so-called ``reflection effect'' becomes important \citep{wil1:90}. This effect accounts for the irradiation of the surface by the other component \citep{prs1:11,prs1:16}. As AzV\,476 is a close binary with two hot O-stars, the reflection effect is modeled with two reflections. The effect of ellipsoidal variability due to tidal interaction, which induces periodic variations in the light curve,  is important in close binary systems with orbital periods on the order of a few days \citep{maz1:08}. However, as none of the binary components are close to filling their Roche lobe (see Table~\ref{tab:PHOEBE_stellar}) and the mass ratio is not extreme ($q_\mathrm{orb}=0.89$) this effect is expected to be negligible \citep{gom1:21}.
    \begin{figure}[t]
        \centering
        \includegraphics[trim= 0.0cm 0cm 0.cm -0.15cm ,clip ,height=0.5\textwidth,angle = -90]{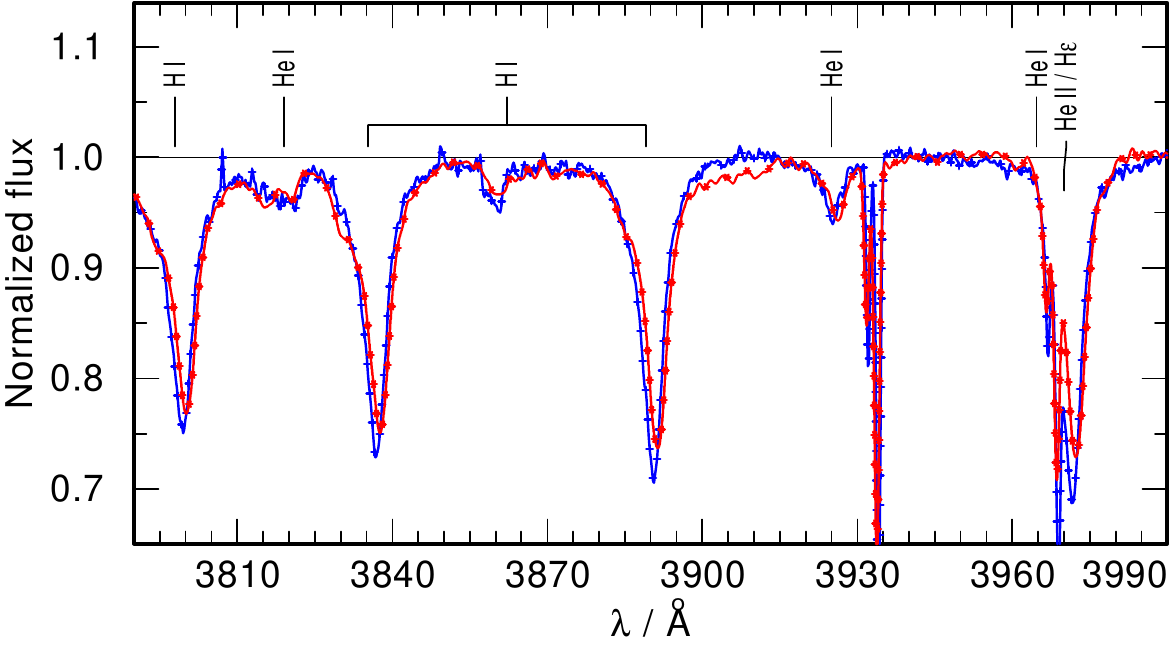}
        \caption{ Same as Fig.\,\ref{fig:rv_shifts}, but now for the region around the \HeI{} $\lambda3819$  and \Hepsilon{} lines. Most of the lines, including \HeI{} $\lambda3926$, are redshifted and associated with the primary, while the \HeI{} $\lambda3819$ line which is associated with the secondary is blueshifted.}
        \label{fig:rv_shifts2}
    \end{figure}

\subsection{Resulting binary parameters}
\label{sec:PHOEBE}

    \begin{table}[tbp]
        \centering
        \caption{Orbital parameters obtained  from the RV and light curves by the PHOEBE code}
        \label{tab:PHEOEB_orbit}
        \begin{tabular}{lcc}
                \hline\hline \rule{0cm}{2.8ex}
                \rule{0cm}{2.8ex}Parameter & \multicolumn{2}{c}{Value}\\ 
                \hline\rule{0cm}{2.8ex}
                $P\,(\mathrm{d})$ & \multicolumn{2}{c}{$9.36665\,\,\,\,(\mathrm{fixed})$}\\ 
                \rule{0cm}{2.8ex}HJD$_0$ & \multicolumn{2}{c}{$7002.7968\,\,\,\,(\mathrm{fixed})$}\\ 
                \rule{0cm}{2.8ex}$e$ & \multicolumn{2}{c}{$0.240^{+0.002}_{-0.002}$}\\ 
                \rule{0cm}{2.8ex}$\omega_0\,(\mathrm{^\circ})$ & \multicolumn{2}{c}{$19^{+2}_{-1}$}\\ 
                \rule{0cm}{2.8ex}$q_\mathrm{orb}$ & \multicolumn{2}{c}{$0.89^{+0.06}_{-0.06}$}\\ \rule{0cm}{2.8ex}$\gamma$ & \multicolumn{2}{c}{$12^{+3}_{-3}$}\\
                \rule{0cm}{2.8ex}$i\,(\mathrm{^\circ})$ & \multicolumn{2}{c}{$77.9^{+0.3}_{-0.3}$}\\ 
                \rule{0cm}{2.8ex}$a\,({R_\odot})$ & \multicolumn{2}{c}{$63^{+2}_{-2}$}\vspace{0.15cm}\\ 
                \hline
        \end{tabular}
        \vspace*{3ex}
        \caption{Stellar parameters obtained from the RV and light curves by the PHOEBE code}
        \label{tab:PHOEBE_stellar}
        \begin{tabular}{lcc}
                \hline\hline \rule{0cm}{2.8ex}
                \rule{0cm}{2.8ex}Parameter & Primary & Secondary\\ 
                \hline
                \rule{0cm}{2.8ex}$T_\mathrm{eff}\,\,[\mathrm{kK}]$ & $42\,\,(\mathrm{input})$ & $32\,\,(\mathrm{input})$\\ 
                \rule{0cm}{2.8ex}$K\,\,[\mathrm{km\,s^{-1}}]$ & $161^{+13}_{-13}$ & $181^{+14}_{-14}$\\ 
                \rule{0cm}{2.8ex}$M_\mathrm{orb}\,\,[{\msun}]$ & $20.2^{+2.0}_{-2.0}$ & $18.0^{+1.8}_{-1.8}$\\ 
                \rule{0cm}{2.8ex}$R\,\,[{\rsun}]$ & $10.7^{+0.4}_{-0.4}$ & $7.0^{+0.4}_{-0.4}$\\ 
                \rule{0cm}{2.8ex}$R_\mathrm{RL}\,\,[{\rsun}]$ & $24.5^{+0.9}_{-0.9}$ & $23.3^{+0.8}_{-0.8}$\\ 
                \rule{0cm}{2.8ex}$\log g\,\,[\si{cm\,s^{-2}}]$ & $3.69^{+0.06}_{-0.06}$ & $4.00^{+0.07}_{-0.07}$\\ 
                \rule{0cm}{2.8ex}$\log L\,\,[\lsun]$ & $5.51^{+0.13}_{-0.13}$ & $4.67^{+0.22}_{-0.22}$\\ 
                \rule{0cm}{2.8ex}$M_\mathrm{bol}\,[\mathrm{mag}]$ & $-8.94^{+0.32}_{-0.32}$ & $-6.83^{+0.56}_{-0.56}$\\ 
                \rule{0cm}{2.8ex}$f_2/f_1\,\,\,\,$({OGLE\,\,I-band})& \multicolumn{2}{c}{$0.31^{+0.05}_{-0.05}$}\vspace{0.15cm}\\ 
                \hline 
        \end{tabular}
    \end{table}

    \begin{table*}[t]
        \centering
        \caption{Summary of the stellar parameters of both stellar components obtained from the different methods.}
        \begin{tabular}{l|cc|cc|cc|cc}\hline \hline \rule{0cm}{2.8ex}%
                                                                          & \multicolumn{2}{c|}{Spectroscopic analysis} & \multicolumn{2}{c|}{Orbital analysis$^{(a)}$}                  &\multicolumn{2}{c|}{Single star evolution$^{(b)}$}       & \multicolumn{2}{c}{Binary evolution$^{(c)}$} \\
                                                         \rule{0cm}{2.2ex}& star 1                 & star 2                  & star 1                     & star 2                  & star 1                   & star 2                   & star 1          & star 2                              \\
                \hline \rule{0cm}{3.4ex}%
                \rule{0cm}{2.8ex}$T_\mathrm{eff}\,\,[\si{kK}]$         &   $42^{+3}_{-3}$       &   $32^{+4}_{-4}$            &   $42$  (fix)              &     $32$ (fix)          &  $40.1^{+1.7}_{-1.5}$   &   $32.5^{+1.8}_{-1.8}$   &   ${42.0}$        &   $31.6$                            \\
                \rule{0cm}{2.8ex}$\log\,g\,\,[\si{cm\,s^{-2}}]$           &$\,\,3.7^{+0.2}_{-0.1}$ &$\,\,4.0^{+0.2}_{-0.2}$  &   $3.69^{+0.06}_{-0.06}$   &   $4.00^{+0.07}_{-0.07}$&   $3.80^{+0.09}_{-0.08}$ &   $4.03^{+0.10}_{-0.11}$ &   $3.56$        &   $4.01$                            \\
                \rule{0cm}{2.8ex}$\log\,L\,\,[\lsun]$                     &  $5.65^{+0.2}_{-0.2}$  &   $4.75^{+0.2}_{-0.2}$  &     $5.51^{+0.13}_{-0.13}$ &   $4.67^{+0.22}_{-0.22}$&   $5.60^{+0.18}_{-0.13}$ &   $4.67^{+0.11}_{-0.14}$ &   $5.55$        &   $4.66$                            \\
                \rule{0cm}{2.8ex}$R\,\,[\rsun]$                           & $12.6^{+1.0}_{-1.0}\,$ & $\,\,7.8^{+2.0}_{-2.0}$ &   $10.7^{+0.4}_{-0.4}$     &   $7.0^{+0.4}_{-0.4}$   &   $12.94^{+2.2}_{-1.7}$  &   $6.57^{+0.8}_{-0.9}$   &   $11.3$        &   $7.2$                             \\
                \rule{0cm}{2.8ex}$M\,\,[\msun]$                           &   $\,\,29^{+17}_{-11}$ &  $\,\,22^{+13}_{-8}$    &     $20.2^{+2.0}_{-2.0}$   &     $18.0^{+1.8}_{-1.8}$&   $39.2^{+8.8}_{-5.8}$   &   $17.4^{+1.4}_{-1.6}$   &   $17.8$        &   $18.2$                            \\
                \rule{0cm}{2.8ex}$M_\mathrm{ini}\,\,[\msun]$              &  ---                   &  ---                    &  ---                       &  ---                     &   $40.2^{+9.1}_{-6.4}$  &   $17.4^{+1.4}_{-1.6}$   &   $33.0$        &   $17.5$                            \\
                \rule{0cm}{2.8ex}$\log \dot{M}\,\,[\msunpyr]$             &  $-6.1^{+0.2}_{-0.2}\,$&   $-8.8^{+0.5}_{-0.5}\,$&  ---                       &  ---                     &   $-6.1^{+0.3}_{-0.3}$  &   $-7.9^{+0.3}_{-0.3}$   &   $-6.4^{(f)}$  &   $-6.46^{(f)}$                     \\
                \rule{0cm}{2.8ex}$\varv_\infty\,\,[\si{km\,s^{-1}}]$      &$2500^{+200}_{-200}$    &$2500$                   &  ---                       &  ---                     &  ---                    &  ---                     &  ---            & ---                                 \\
                \rule{0cm}{2.8ex}$\varv \sin i\,\,[\si{km\,s^{-1}}]$      &   $\,\,\,140^{(d)}$    &   $\,\,\,\,\,425^{(d)}$ &  ---                       &  ---                     &   $150^{+26}_{-28}$     &   $410^{+60}_{-40}$      &   $96$          &   $575$                             \\
                \rule{0cm}{2.8ex}$X_\mathrm{H}\,\,(\mathrm{by\,\,mass})$  &$0.73$                  &$0.73$                   &  ---                       &  ---                     &  $0.74^{+0.0}_{-0.1}$   &   $0.74^{+0.0}_{-0.1}$   &   $0.47$        &   $0.74$                            \\
                \rule{0cm}{2.8ex}$X_\mathrm{C}/10^{-5}\,\,(\mathrm{by\,\,mass})$&$\,\,\,2^{+2}_{-1}$&$\,\,\,21^{(e)}$        &  ---                       &  ---                     &   $27^{+2}_{-5}$        &   $7^{+5}_{-3}$          &   $3$           &   $20$                              \\
                \rule{0cm}{2.8ex}$X_\mathrm{N}/10^{-5}\,\,(\mathrm{by\,\,mass})$&$\,\,\,45^{+5}_{-10}$&$\,\,\,\,\,3^{(e)}$   &  ---                       &  ---                     &   $\,\,\,6^{+14}_{-6}$  &   $61^{+40}_{-30}$       &   $91$          &   $7$                               \\                
                \rule{0cm}{2.8ex}$X_\mathrm{O}/10^{-5}\,\,(\mathrm{by\,\,mass})$&$\,\,\,80^{+10}_{-20}$&$\,110^{(e)}$        &  ---                       &  ---                     &  $91^{+4}_{-14}$        &   $70^{+7}_{-37}$        &   $35$          &   $110$                             \\
                \rule{0cm}{2.8ex}$\log\,Q_\ion{H}\,\,[\si{\,s^{-1}}]$     &   $49.43$              &   $47.88$               &  ---                       &  ---                     &    ---                  &  ---                     &  $49.34^{(g)}$  &  $47.72^{(g)}$                      \\                
                \rule{0cm}{2.8ex}$\log\,Q_{\HeI{}}\,\,[\si{\,s^{-1}}]$    &   $48.72$              &   $45.83$               &  ---                       &  ---                     &    ---                  &  ---                     &  $48.62^{(g)}$  &   $45.76^{(g)}$                     \\                
                \rule{0cm}{2.8ex}$\log\,Q_{\HeII{}}\,\,[\si{\,s^{-1}}]$   &   $40.80$              &   $41.38$               &  ---                       &  ---                     &    ---                  &  ---                     &  $43.63^{(g)}$  &   $35.75^{(g)}$                     \\
                \rule{0cm}{2.8ex}$\mathrm{age}\,\,[\si{Myr}]$             & ---                    &  ---                    &  ---                       &  ---                     &   $2.8^{+0.5}_{-0.4}$   &   $5.9^{+1.5}_{-1.5}$    &  $6.0$          &   $6.0$                             \\
                        \rule{0cm}{-1.2ex}                                        &                        &                         &                            &                          &                         &                          &                 &                                     \\
                        \hline
        \end{tabular}
        \rule{0cm}{2.8ex}%
        \begin{minipage}{0.95\linewidth}
            \ignorespaces
            $^{(a)}$ Results from the PHOEBE code, see Sect.~\ref{sec:PHOEBE}. $^{(b)}$ Results from the BONNSAI tool, see Sect.~\ref{sec:single_stars}. $^{(c)}$ Models calculated with MESA, see Sect.~\ref{sec:binary_stars}. No uncertainties are given, as we did no in-depth analysis. $^{(d)}$ Obtained with the \verb|iacob| broad tool. $^{(e)}$ Initial CNO abundances scaled to SMC metallicity; for more details see Sect.~\ref{sec:spectral_fitting}. $^{(f)}$ According to the mass-loss recipe implemented in our evolutionary models. $^{(g)}$ Values taken from the spectroscopic model that is calculated with the parameters of the binary evolutionary models.%
        \end{minipage}
        \label{tab:stellar_parameters_summary}
    \end{table*}
    
    The best fitting model light curve and RV fit obtained with the PHOEBE code are shown in Fig.~\ref{fig:LC_and_RV_fit}. The corresponding orbital parameters are listed in Table~\ref{tab:PHEOEB_orbit} and the stellar parameters of both components in Table~\ref{tab:PHOEBE_stellar}. 
    The orbital solution yields similar masses for both components, while the light curve fit reveals different $T_\mathrm{eff}$ and $L$. This indicates that there was a prior mass-transfer phase that has stripped away most of the primary's envelope, such that now has similar mass to its companion. 
    From the light-curve fit, the fundamental stellar parameters --stellar radius and the surface gravity--- are determined, giving us the opportunity to cross-check the spectral analysis. Furthermore, the PHOEBE code calculates the light ratio of the binary components in the observed band outside conjunctions, which is compared to the light ratio obtained from the spectroscopic analysis in Sect.\,\ref{sec:spectral_analyis}.

\section{Spectral analysis}
\label{sec:spectral_analyis}

\subsection{Method: Spectral modeling with PoWR}
   \label{sec:spec_analysis}
   
   Synthetic spectra for both stellar components were calculated with the  Potsdam Wolf-Rayet (PoWR) model atmosphere code \citep{gra1:02,ham1:04}. In the following, we briefly describe the code. For further details, we refer to \cite{gra1:02}, \cite{ham1:03}, \cite{tod2:15}, and \citet{san2:15}. 
   
   The PoWR code models stellar atmospheres and winds permitting departures from local thermodynamic equilibrium (non-LTE). The code has been widely applied to hot stars at various metallicities \citep[e.g.,][]{hai1:14,hai1:15,osk1:11,rei1:14,she1:15}. The models are assumed to be spherically symmetric, stationary, and in radiative equilibrium.  Equations of statistical equilibrium are solved in turn with the radiative transfer in the co-moving frame. Consistency is achieved iteratively using the ``accelerated lambda operator'' technique. This yields the population numbers within the photosphere and wind.
   
   The emergent spectrum in the observer's frame is calculated with the formal integral in which the Doppler velocity is split into the depth-dependent thermal velocity and a ``microturbulence velocity'' $\zeta(r)$. The microturbulence grows from its photospheric value $\zeta_\mathrm{ph}=\SI{10}{km\,s^{-1}}$ linearly with the wind velocity up to $0.1\varv_\infty$. 
   
   By comparison of the synthetic spectrum with the observations, it is possible to determine the main stellar parameters. In addition to the chemical composition, the main stellar parameters that specify the model atmosphere are the luminosity $L$, stellar temperature $T_*$, surface gravity $g_*$, wind mass-loss rate $\dot{M}$, and the wind terminal wind velocity $\varv_\infty$. The stellar temperature is defined as the effective temperature referring to the stellar radius $R_*$ by the Stefan-Boltzmann equation $L=4\pi \sigma R_*^2 T_*^4$. The stellar radius is defined at the Rosseland mean optical depth of $\tau = 20$. The differences between $T_*$ and the effective temperature $T_\mathrm{eff}$ (referring to the radius where the optical depth $\tau=2/3$) are negligibly small, as the winds of OB-type stars are optically thin.
   
   In the subsonic regions of the stellar atmosphere, the velocity law is calculated such that the density, related by the equation of continuity, approaches the hydrostatic stratification. The hydrostatic equation consistently accounts for the radiation pressure \citep{san2:15}. In the supersonic regions, it is assumed that the wind velocity field can be described by a so-called double $\beta$-law as was first introduced by \citet{hil1:99}: 
   \begin{equation}
       \varv(r) = \varv_\infty\left[\left(1-f\right)\left(1-\dfrac{r_0}{r}\right)^{\beta_1}+f\,\left(1-\dfrac{r_1}{r}\right)^{\beta_2}\right].
   \end{equation}
   In this work, we assume $\beta_1=0.8$, a typical value for O-stars, $\beta_2=4$, and a contribution of $f=0.4$ of the second $\beta$-term. $r_0$ and $r_1$ are close to the stellar radius $R_*$ and are determined such that the quasi-hydrostatic part and the wind are smoothly connected.
   This choice of the wind velocity law leads to better agreement between the \Halpha{} absorption line and the \CIV{} P\,Cygni line profile than the classical $\beta$-law \citep{cas1:75}.
   
   Inhomogeneities within the wind are accounted for as optically thin clumps (``microclumping'') which are specified by the ``clumping factor'' $D,$ which describes by how much the density within the clumps is enhanced compared to a homogeneous wind with the same mass-loss rate \citep{ham2:98}. In our analysis, we use a depth-dependent clumping which starts at the sonic point and increases outward until a clumping factor of $D=20$ is reached at a radius of $R_\mathrm{D}=7\,R_*$ for the primary and at a  radius $R_\mathrm{D}=10\,R_*$ for the secondary. The smaller radius at which the clumping factor is reached in the primary's wind profile is needed to model the observed \OV{} $\lambda1371$ line properly. 
    
    \begin{figure*}[ht]
        \centering
        \includegraphics[trim= 0.cm 0.cm 0cm 1.1cm ,clip ,height=\textwidth,angle = -90]{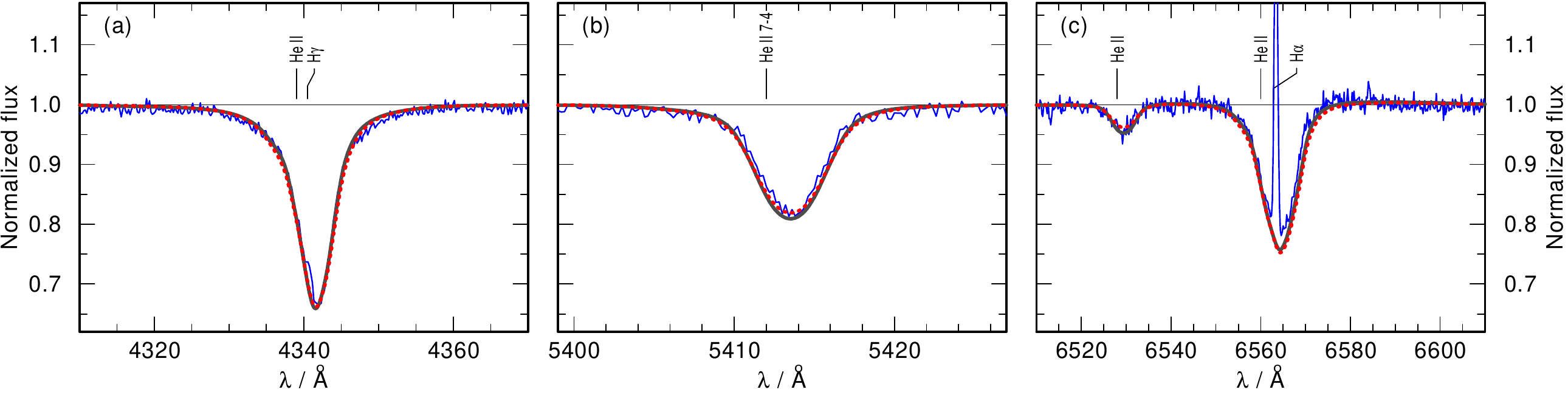}
        \caption{Selected regions of the X-SHOOTER spectrum (ID 5 in Table~\ref{tab:spectroscopy_and_RV}) compared to different synthetic spectra. The observed spectrum, corrected for the velocity of the SMC and the barycentric motion, is shown as a solid blue line. The red dotted line is our best-fitting model which consists of the primary with ${T_\mathrm{eff,\,1}=\SI{42}{kK}}$ and $\log g_{*,\,1} = 3.7$ and the secondary with ${T_\mathrm{eff,\,2}=\SI{32}{kK}}$ and $\log g_{*,\,1} = 4.0$. The gray solid line is again a combined synthetic spectrum of the primary and secondary, but this time the surface gravity of the primary has been reduced to $\log g_{*,\,1} = 3.6$ (and the temperature and mass-loss rate are slightly adjusted so that the spectrum matches the observations) while the parameters of the secondary are kept fixed.  The line identification marks correspond to the wavelengths in the rest frame. Panel (a) shows the \Hgamma{} line, in which the red wing is dominated by the primary and the blue wing by the secondary. Panel (b) shows the \HeII{}~$\lambda5412$ absorption line, which is dominated by the primary and is  sensitive to surface gravity. Panel (c) shows the region of \HeII{}~$\lambda6528$ and \Halpha{}.  While the \Halpha{} wings are only barely affected by the change in $\log g_{*,\,1}$, the \HeII{} $\lambda6528$ line is more sensitive to it.}
        \label{fig:balmer}
    \end{figure*}

   The PoWR model atmospheres used here account for detailed model atoms of H, He, C, N, O, Mg, Si, P, and S. The iron group elements Sc, Ti, V, Cr, Mn, Fe, Co, and Ni are combined to one generic element ``G'' with solar abundance ratios, and  treated in a superlevel approach \citep{gra1:02}. 
   The abundances of Si, Mg, and Fe are based on \citet[their table 17]{hun1:07} and \citet[their table~9]{tru1:07}. For the remaining elements, we divide the solar abundances of \citet[their table~1]{asp1:05} by seven to match the previously mentioned  metallicity of the SMC. This yields the following mass fractions:
   ${X_\mathrm{H}=0.73}$, ${X_\mathrm{Si}=\num{1.3e-4}}$, ${X_\mathrm{Mg}=\num{9.9e-5}}$, ${X_\mathrm{P}=\num{8.32e-7}}$, ${X_\mathrm{S}=\num{4.42e-5}}$, and ${X_\mathrm{G}=\num{3.52e-4}}$. 
   The CNO individual abundances of both stellar components are not fixed but derived from the analysis. The complement mass fraction to unity is $X_{\rm He}$.
   
\begin{figure*}[htpb]
        \centering
        \includegraphics[width=\textwidth]{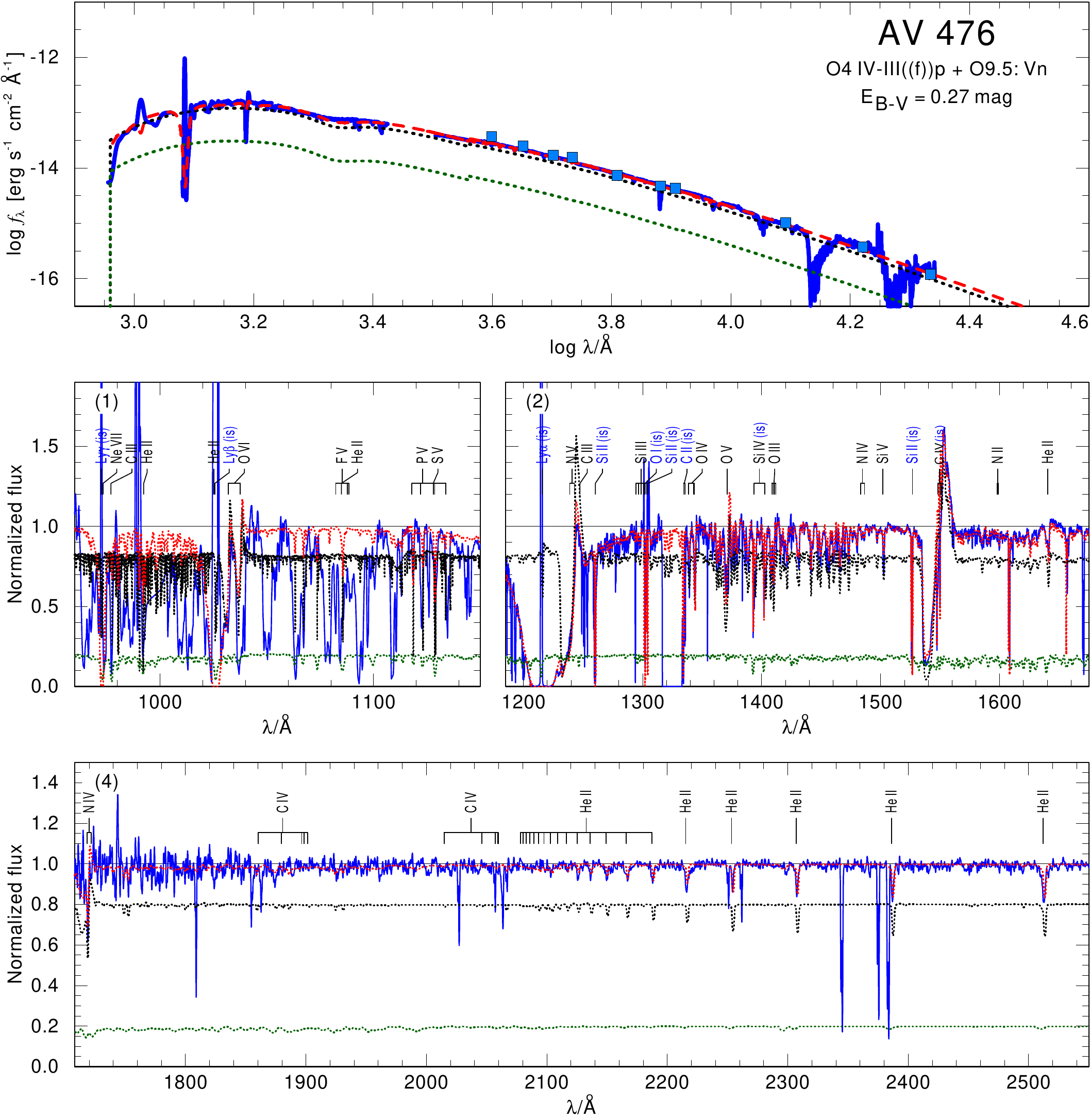}
        \caption{Spectral fit for AzV\,476. \textit{Top panel:} SED. Flux-calibrated observations (blue lines) are the FUSE (1), COS (2), STIS (4), and X-SHOOTER (5) spectra as listed in Table~\ref{tab:spectroscopy_and_RV}. The light blue squares are the photometric UBVRIJHK data as listed in Table~\ref{tab:photometry}. The model SED composed of both stellar components is shown as a dashed red line, while the individual SEDs of the primary and the secondary are shown as dotted black and green lines, respectively. \textit{Lower panels:} Normalized FUSE (1), COS (2), and STIS (4) spectra. The number in the upper left corner corresponds to the ID given to a specific spectrum as listed in Table~\ref{tab:spectroscopy_and_RV}. The line styles are the same as in the top panel. The synthetic spectra are calculated with the model parameters compiled in Table~\ref{tab:stellar_parameters_summary} (``Spectroscopy'' columns).}
        \label{fig:spectral_fit}
    \end{figure*}

   We determined the color excess $E_\mathrm{B-V}$ and the luminosity $L$ of the binary components by fitting the composite SED to photometry (top panel in Fig.~\ref{fig:spectral_fit}). Reddening is modeled as a combined effect of the Galactic foreground, for which we adopt the reddening law of \citet{sea1:79} with $E_\mathrm{B-V} = \SI{0.03}{mag}$, and the reddening law of \citet{bou1:85} for the SMC. 
   
   We use the \verb|iacob-broad| tool \citep{sim1:14} in combination with the high $\mathrm{S/N}$ X-SHOOTER spectrum ($\mathrm{S/N}\sim100$) to determine the rotation rates of the primary and the secondary. The helium lines are potentially pressure broadened and thus are not optimally suited for rotation broadening measurements. Therefore, 
   for the primary, we use the few metal lines that are visible in the optical spectrum, namely  the \NIV{}\,$\lambda3478$ and the \OIV{}\,$\lambda3403$ absorption lines and the \NIV{}\,$\lambda4058$ emission line.  We determine  a projected rotation rate of ${\varv_1 \sin i = \SI{140}{km\,s^{-1}}}$ for the primary. The secondary only contributes to the \HeI{} lines. Fitting the \HeI{}\,$\lambda4387$ and \HeI{}\,$\lambda4471$ absorption lines ---in which the contribution of the primary is smallest--- yields a rotation rate of ${\varv_2 \sin i = \SI{425}{km\,s^{-1}}}$ for the secondary.

    \begin{figure}[ht]
        \centering
        \includegraphics[width=0.45\textwidth,angle = -90]{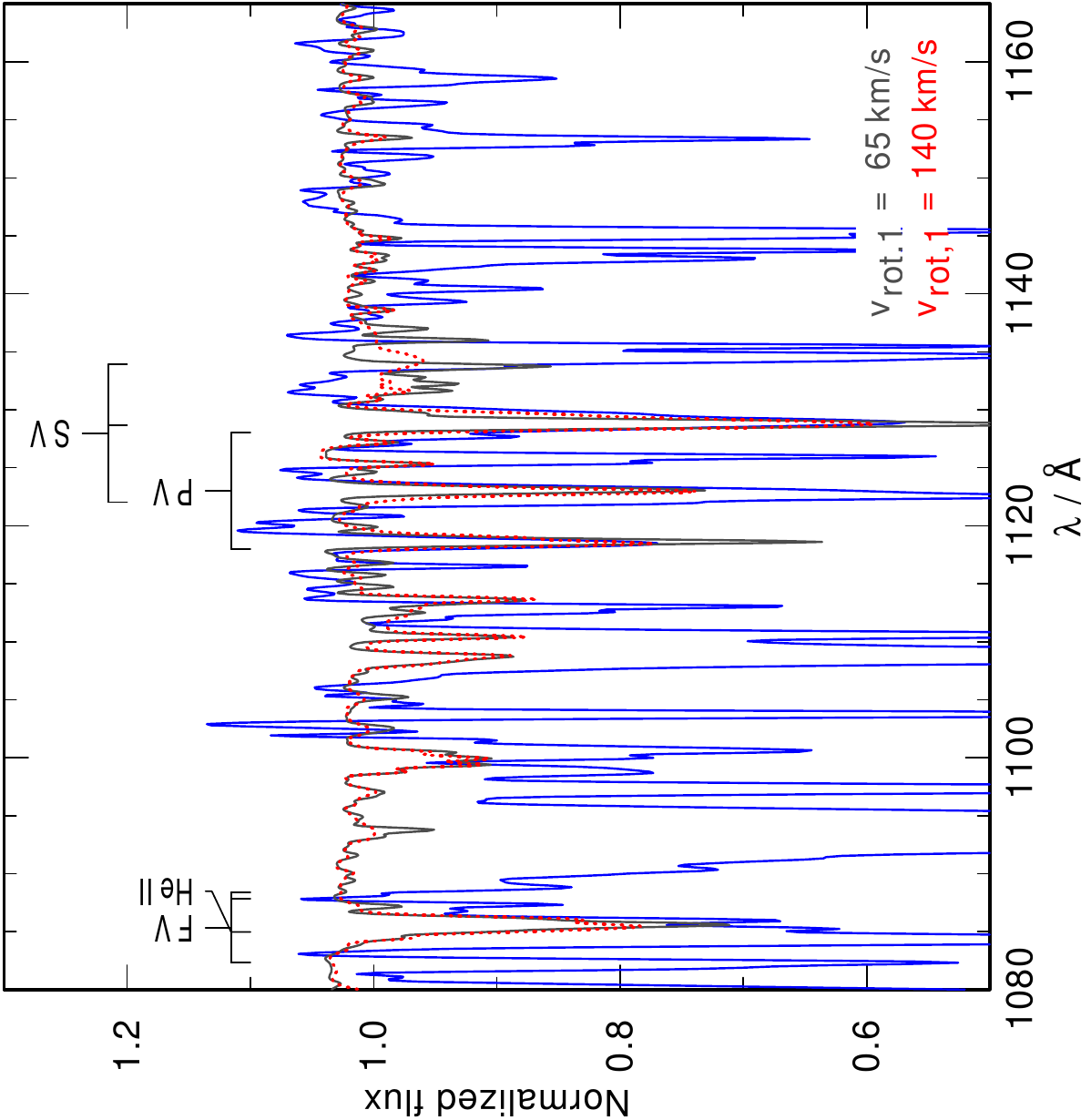}
        \caption{Part of the observed  FUSE 
        spectrum (ID 1 in Table~\ref{tab:spectroscopy_and_RV}) covering the wavelength range of the LiF2 channel -- the same as used by \citet{pen1:09} to determine the projected rotational velocity.
        The observed spectrum, corrected for the velocity of the SMC, is shown as a solid blue line.  
        The red dotted line is our best-fitting model  with the primary's projected rotational velocity  ${\varv_1 \sin i = \SI{140}{km\,s^{-1}}}$. The gray solid line is another model where the primary's projected rotational velocity is reduced to ${\varv_1 \sin i = \SI{140}{km\,s^{-1}}}$. The line identification marks correspond to the wavelengths in the rest frame.}
        \label{fig:vrot_FUSE}
    \end{figure}
   
   Previously, \citet{pen1:09} determined the projected rotational velocity of the primary to ${\varv_1 \sin i = \SI{65}{km\,s^{-1}}}$ using a cross-correlation of the FUSE spectrum in the far-UV with a template spectrum. We calculate tailored spectral models with the respective projected rotational velocities and find that also the FUSE spectrum is better reproduced when using ${\varv_1 \sin i = \SI{140}{km\,s^{-1}}}$ (see Fig.~\ref{fig:vrot_FUSE}). We assume that the template spectrum used in \citet{pen1:09} might not have been perfectly calibrated for our target and that the binary nature leads to additional uncertainties.
   
\subsection{Resulting spectroscopic parameters}
\label{sec:spectral_fitting}
\subsubsection{Temperature and surface gravity of the primary}
    AzV\,476 was previously classified as O2-3\,V plus a somewhat later O-type companion.
    This implies that nitrogen lines are expected in absorption and emission in the primary spectrum, while \HeI{} absorption lines should be present in the secondary spectrum.
    
    Indeed, we find that the \NIV{}~$\lambda4057$ (hereafter \NIV{}) emission line,\, the \NIV{} absorption lines at $\lambda\lambda\,\,\SI{3463}{\AA}$, $\SI{3478}{\AA}$, $\SI{3483}{\AA,}$ and $\SI{3485}{\AA,}$ and the \HeII{}~$\lambda6528$ absorption can be entirely assigned to the primary. We observe only marginal \NIII{}~$\lambda\lambda4634, 4640$ (hereafter \NIII{}) emission and \NV{}~$\lambda\lambda4603,4619$ (hereafter \NV{}) absorption; see Fig.~\ref{fig:nitro2}. Thus, the optical spectrum shows nitrogen only as \NIV{}, while \NIII{} as well as \NV{} are virtually absent. This restricts the temperature of the primary to a narrow range \citep{riv1:12}. We also find that the \OIV{} multiplets around $\SI{3400}{\AA}$ are purely associated with the primary. Because these lines are highly contaminated by ISM absorption lines of \ion{Ti}{ii}~$\lambda3385$ and \ion{Co}{i} $\lambda3414,$ they are only used as a crosscheck of the oxygen abundance applied in our spectral model. A more detailed description of this line complex is given in Appendix~\ref{app:OIV}.
    
    The secondary does not contribute noticeably to these weak metal lines. However, the secondary strongly dominates the \HeI{} absorption lines at $\lambda\lambda$ $\SI{3819}{\AA}$, $\SI{4387}{\AA,}$ and $\SI{4471}{\AA,}$ indicating a lower effective temperature. The primary star also contributes to these lines, giving additional constraints on the temperature of the  primary.
    
    The surface gravity $g_{*,\,1}$ of the primary is determined by fitting the wings of the  Balmer lines using those UVES spectra with the least wavy patterns and highest RV shifts as well as the X-SHOOTER spectrum, which has a higher $\mathrm{S/N}$. Because both stars contribute to the Balmer lines, the \HeII{}~$\lambda5412$ absorption line that is mostly originating from the primary and also sensitive to changes in surface gravity is used. Unfortunately, this line is only contained in the X-SHOOTER spectrum. Surface gravity affects the density structure and therefore the ionization balance of, for example, nitrogen and helium. As the temperature is adjusted such that the nitrogen lines are reproduced, changes in the surface gravity will not only affect the wings of the \HeII{} lines but also change  the depth of the \HeII{} absorption lines. This gives the opportunity to use the \HeII{}~$\lambda6528$ absorption line in the UVES spectra as second criterion for the surface gravity estimate in the primary.
    
    We tested different  temperature values in the range ${T_\mathrm{eff,\,1}=\SIrange{39}{50}{kK}}$ and surface gravities in the range   ${\log(g_{*,\,1}/(\si{cm\,s ^{-2}}))=\numrange{3.5}{4.1}}$ and found that a temperature of ${T_\mathrm{eff,\,1}=\SI{42\pm3}{kK}}$ and a surface gravity of $\log(g_{*,\,1}/(\si{cm\,s ^{-2}}))=3.7\pm0.2$ are most suitable for reproducing the primary's spectrum. The given error margins take into account that the measurements of temperature and surface gravity are not independent. 
    
    Different regions of the X-SHOOTER spectrum that are used to determine the surface gravity are shown in Fig.~\ref{fig:balmer}. The accuracy in surface gravity highly depends on the calibration of the spectrum. One can see that \Hgamma{} and \HeII{}~$\lambda5412$ line fits show preference to a surface gravity for the primary of $\log(g_{*,\,1}/(\si{cm\,s ^{-2}}))=3.7$, while \Halpha{} and \HeII{}~$\lambda6528$ indicate that $\log(g_{*,\,1}/(\si{cm\,s ^{-2}}))=3.6$ might be more suitable. However, from fitting the UVES data with the highest RV shifts we find that a surface gravity for the primary of $\log(g_{*,\,1}/(\si{cm\,s ^{-2}}))=3.7$ yields the best fit.
    
\subsubsection{Temperature and surface gravity of the secondary}
    
    It is more difficult to determine the temperature of the  secondary, as only a few \HeI{} lines are associated with the secondary and all these line have a contribution from the primary. However, from the spectra with the highest RV shifts (e.g., UVES spectrum with ID 10), we find that the secondary does not contribute to the \HeII{} $\lambda4200$ line, and therefore we have an additional constraint that can be used as an upper limit on the temperature of the 
secondary. In addition to the limited number of lines that are associated with the secondary, the ionization balance that changes depending on the surface gravity introduces another uncertainty in our temperature estimation.
    
    Therefore, we first adjust $g_{*,\,2}$ of the secondary such that the wings of the Balmer lines are well reproduced while keeping the surface gravity and temperature of the primary fixed. As in the case of the primary, in order to distinguish the  contributions of the binary components to the Balmer wings, we employ the UVES spectra in the phase with the largest RV shifts. This gives additional information and a surface gravity of ${\log(g_{*,\,2}/(\si{cm\,s ^{-2}}))=4.0}$ can be determined. Finally, the temperature of the secondary is adjusted such that the \HeI{} absorption lines at $\lambda\lambda$ $\SI{3819}{\AA}$, $\SI{4387}{\AA,}$ and $\SI{4471}{\AA}$ match the observation. Following this procedure, the spectrum of the secondary is best reproduced with a temperature of ${T_\mathrm{eff,\,2}=\SI{32}{kK}}$.  This method is accurate to $\Delta T_\mathrm{eff,\,2}=\SI{\pm4}{kK}$ in temperature and $\Delta \log(g_{*,\,2})=\pm0.2$ for the surface gravity. The obtained values of the surface gravity of the primary and the secondary are in agreement with those obtained from the orbital analysis (see Sect.\,\ref{sec:orbit_analyis}).
    
\subsubsection{Luminosity}
    To estimate the light ratio, the luminosities of both stars are adjusted such that the shapes of the optical \HeII{} lines match the observations. This is an iterative process done simultaneously with the temperature and surface gravity estimate. The total luminosity of the system is calibrated such that the calculated visual magnitude of the synthetic flux matches the observed one. This results in luminosities of $\log(L_1/\lsun) = 5.65$ for the primary and $\log(L_2/\lsun) = 4.75$ for the secondary. This estimate is sensitive to  synthetic spectra, light ratio, and  distance modulus. Therefore, we estimate that our measurements of the luminosity of each stellar component are accurate to $\Delta \log(L/\lsun) =\pm0.2$. 

    Taking these uncertainties into account, the model flux ratio in the OGLE I-band yields  $f_\mathrm{2}/f_\mathrm{1}= 0.26\pm0.1$ which is in agreement with the flux ratio derived using PHOEBE ${f_\mathrm{2}/f_\mathrm{1}= 0.31\pm0.05}$ (see Sect.~\ref{sec:orbit_analyis}).

    \begin{figure}[t]
        \centering
        \includegraphics[height=0.5\textwidth,angle = -90]{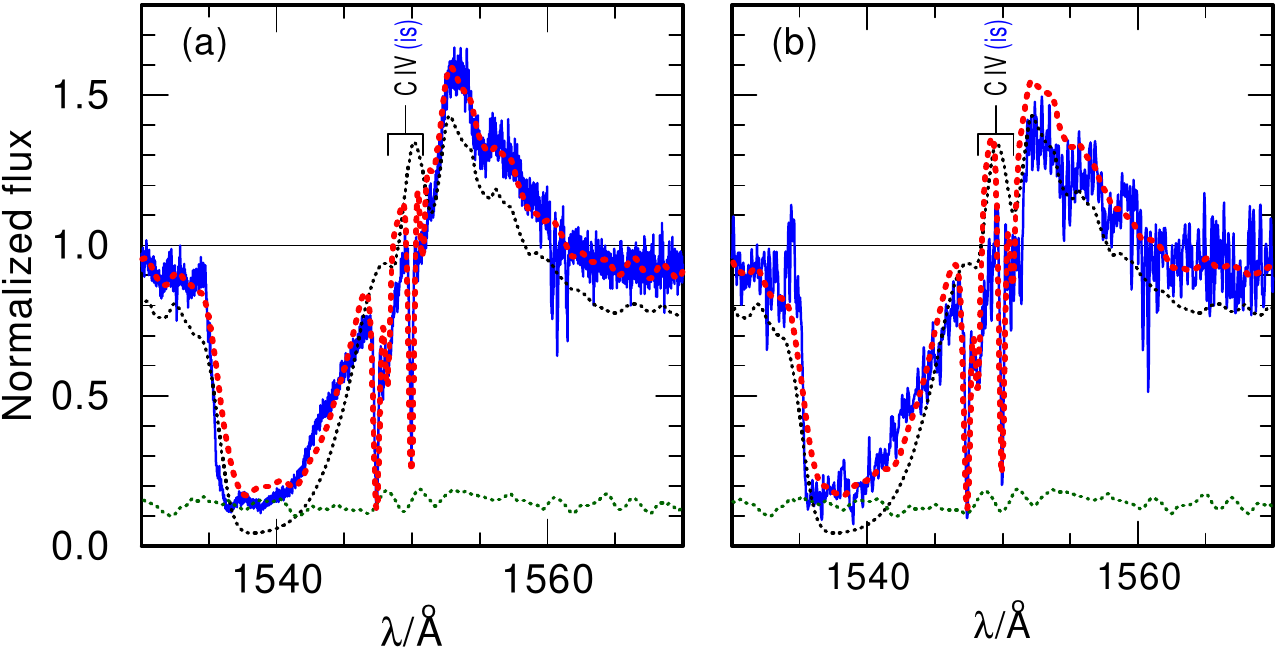}
        \caption{Observed (blue) and synthetic (red) \CIV{} resonance doublet of our final model. The individual primary and secondary spectra are shown as dotted black and green lines, respectively. The sharp interstellar absorptions arise in the Galactic foreground and in the SMC, and are also modeled with their respective RV shifts. (a) COS observation (ID 2 in Table~\ref{tab:spectroscopy_and_RV}) (b) STIS observation (ID 3 in Table~\ref{tab:spectroscopy_and_RV}). }
        \label{fig:civ}
    \end{figure}
    
    \begin{figure*}[ht]
        \centering
        \includegraphics[trim= 0.cm 0.cm 0cm 1.5cm ,clip ,height=\textwidth,angle = -90]{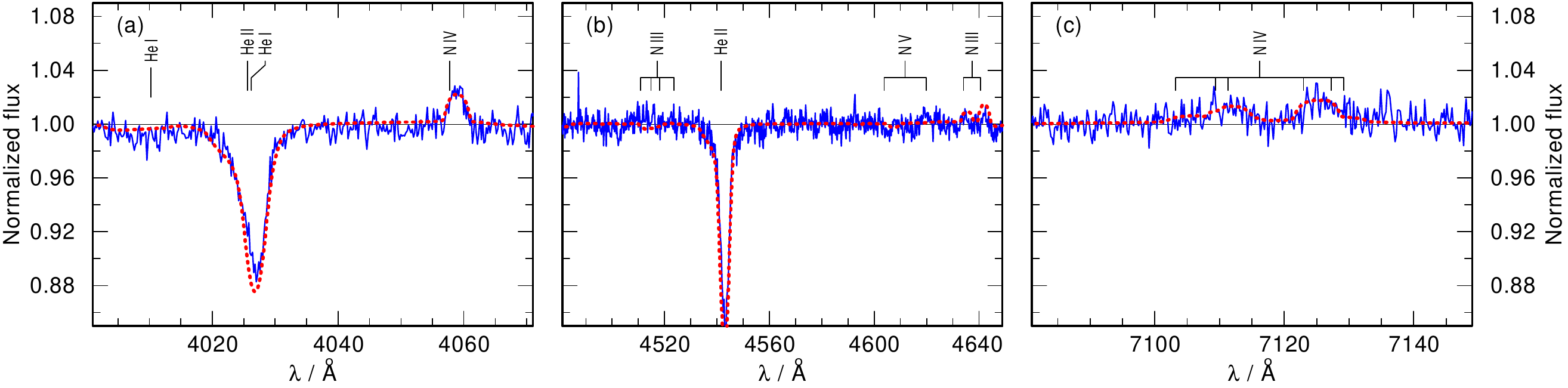}
        \caption{Observed X-SHOOTER spectrum (ID 5 in Table~\ref{tab:spectroscopy_and_RV}) and synthetic spectra of selected regions that show nitrogen emission lines. The observed spectrum is shown as a solid blue line, while the dotted red line is our best-fitting model. The observed spectrum is corrected for the velocity of the SMC and the barycentric motion. The line identification marks correspond to the wavelengths in the rest frame.  Panel (a) shows the area around the \NIV{} $\lambda4057$. Panel (b) shows the region of \NIII{} $\lambda\lambda4511-4523$, \NV{} $\lambda\lambda4604,4620,$ and \NIII{} $\lambda\lambda4634,4641$. The \NIII{} and \NV{} lines are not observable in the spectrum. Panel (c) shows the region of the \NIV{} $\lambda\lambda7103-7129$ complex.}
        \label{fig:nitro}
    \end{figure*}

\subsubsection{Mass-loss rates}
    After determining the temperatures, surface gravities, and luminosities of both binary components, we proceed to measuring the properties of their stellar winds. The \Halpha{} line in the optical as well as the \CIV{} resonance line in the UV are used as the main diagnostic tools for measuring the mass-loss rate of the primary. In addition, we also pay attention to the appearance of the optical \HeI{}\,$\lambda4686$ and the UV \HeII{}\,$\lambda1640$ line; see Fig\,\ref{fig:wind}. When calibrating the  mass-loss rate of the primary, the best fit is archived with a mass-loss rate of  $\log(\dot{M}_1/(\msunpyr))=-6.1\pm0.2$ and a terminal wind velocity of $\varv_{\infty,\,1} = \SI{2500}{km\,s^{-1}}$. 
    
    We have two spectra covering the \CIV{} resonance line at different phases, the COS and the STIS spectrum (ID 2 and 3 in Table~\ref{tab:spectroscopy_and_RV}), depicted in Fig.~\ref{fig:civ}. The \CIV{} P\,Cygni line profile in both spectra does not show any direct indications for the contribution from the secondary such as a double emission peak or a step in the absorption trough. The  \CIV{} resonance line in the STIS spectrum shows slightly weaker emission, which is most likely due to the sensitivity of the different instruments or weak wind variability. We use the bottom part of the P\,Cygni profile of \CIV{} resonance line to confirm the continuum contribution of the secondary in the UV and therefore the used light ratio ($L_2/L_1$).
    
    Apparently the secondary does not contribute to the wind lines. Its mass-loss rate can therefore only be limited to ${\log(\dot{M}_2/(\msunpyr))\leq-8.8\pm0.5}$. For higher mass-loss rates, we find that the secondary would noticeably contribute to the  \CIV{} resonance line in the UV. Due to the lack of indications of the secondary's wind we adopt the terminal velocity of the primary, $\varv_{\infty,\,2} = \SI{2500}{km\,s^{-1}}$. 
    
    \begin{figure*}[htpb]
        \centering
        \includegraphics[width=\textwidth]{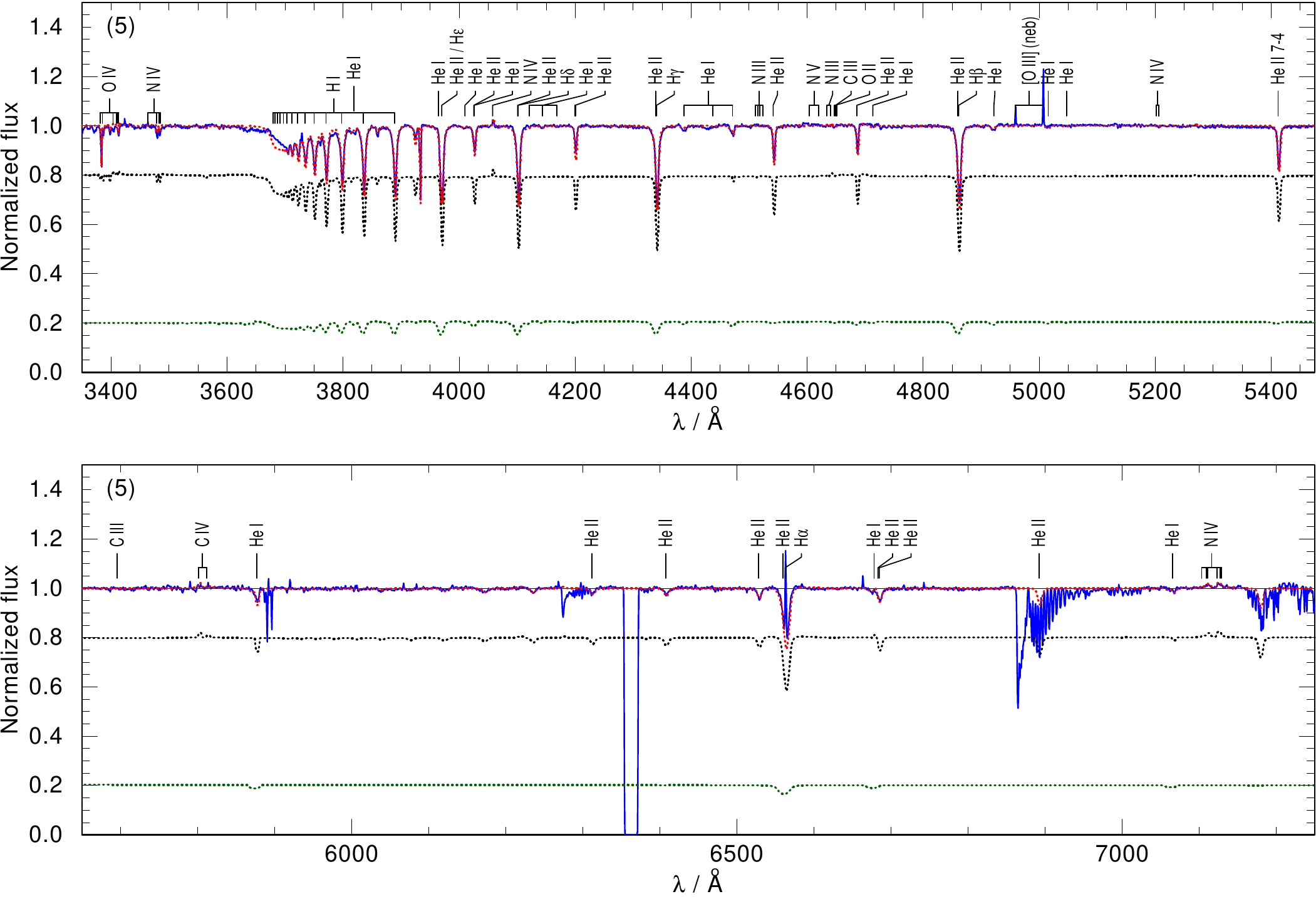}
        \caption{Same as Fig.\,\ref{fig:spectral_fit}, but now the normalized optical X-SHOOTER spectra (ID 5 in Table~\ref{tab:spectroscopy_and_RV}) are shown.}
        \label{fig:spectral_fit2}
    \end{figure*}

\subsubsection{CNO surface abundances}
    As already explained in Sect.~\ref{sec:spec_analysis}, the abundances of hydrogen and the iron group elements are fixed, while the CNO abundances of each stellar component are adjusted during spectral modeling.
    The nitrogen abundance is adjusted to find good agreement with the optical \NIV{} absorption lines at wavelengths $\SI{3463}{\AA}$, $\SI{3478}{\AA}$, and $\SI{3483}{\AA}$, and the \NIV{}~$\lambda4057$ and  \NIV{}~${\lambda\lambda7103-7129}$ emission lines,
    as well as the UV \NV{}~$\lambda\lambda1238, 1242$ resonance doublet. The best agreement is achieved with a nitrogen mass fraction of ${X_\mathrm{N}= (45^{+5}_{-10})\times\num{e-5}}$. The nitrogen lines seen in the optical are displayed in Fig.~\ref{fig:nitro}.
    
    A carbon abundance of ${X_\mathrm{C}= (2^{+2}_{-1})\times\num{e-5}}$ reproduces the optical \CIV{}~$\lambda\lambda5801, 5812$ lines best while maintaining the fit of the \CIV{} resonance doublet, at least for mass-loss rates that do not conflict with other wind features. For lower abundances, the latter is no longer saturated. The oxygen abundance of the primary is calibrated with the oxygen absorption lines in the UV and the optical \OIV{} multiplets around $\SI{3400}{\AA}$,
    resulting in ${X_\mathrm{O}= (80^{+10}_{-20})\times\num{e-5}}$. 
    
    There are no strong CNO lines in the secondary spectrum, and therefore we adopt initial CNO abundances scaled to the metallicity of the SMC ($Z_\mathrm{SMC}=1/7\,\zsun$). Alternatively,  one could assume that the surface abundances in the secondary are close to those in the primary as the accreted material is polluting the surface of the 
secondary. We explored both assumptions but could not find diagnostic lines that would allow us to pin down the composition of the atmosphere and wind of the  secondary. We are only able to fix upper limits for the CNO abundances such that these elements would not produce features that are inconsistent with the observations:
    $X_\mathrm{N}\lesssim50\times\num{e-5}$, $X_\mathrm{C}\lesssim21\times\num{e-5}$ and $X_\mathrm{O}\lesssim110\times\num{e-5}$.

\subsubsection{Required X-ray flux}    

    Similar to other O-type stars, AzV\,476 shows a notoriously strong P\,Cygni profile in the 
    \OVI{} $\lambda\lambda1032,\,1038$ resonance doublet. However, wind models do not predict a sufficient population of this high ion without the inclusion of additional physical processes. This phenomenon, termed {``super-ionization''}, was described by \citet{cas1:79} and interpreted as evidence for the presence of an X-ray field in stellar atmospheres.
    
    In order to model the \OVI{} doublet in the observed spectrum of AzV\,476, we add a hot plasma component. 
    The X-ray-emitting plasma has an adopted temperature of $T_\mathrm{X}=\SI{3}{MK}$ and is distributed throughout the wind outside a radius of $R_\mathrm{X}=1.1\,R_*$. Its constant filling factor is a free parameter; the best reproduction of the observation is achieved for a model with an emergent X-ray luminosity of ${\log\,L_\mathrm{X}=\SI{31.4}{erg\,s^{-1}}}$. 
    
    Current X-ray telescopes are not sensitive enough to detect individual O stars in the SMC. Our final model predicts an X-ray flux at earth of $\SI{1.4e-17}{erg\,s^{-1}\,cm^{-2}\,\AA^{-1}}$ integrated over the X-ray band ($\approx 6 - 60$\,\AA). The region on the sky where AzV\,476 is located was observed by both {\em XMM-Newton} and {\em Chandra} X-ray observatories. However, despite the detection of AzV\,476 in the UV by the {\em XMM-Newton} optical monitor, the star is not detected in X-rays. Hence, we set the upper limit on its X-ray
    flux at the median flux of detected sources in the XMMSSC - XMM-Newton Serendipitous Source Catalog (4XMM-DR10 Version); this is $\SI{5e-15}{erg\,s^{-1}\,cm^{-2}}\,\AA^{-1}$ , which  is far from being sufficiently sensitive for our predicted source. 
    
    The \OVI{} resonance doublet is not the only feature that is sensitive to X-rays. Other ions which might be populated by Auger ionization are \NV{} and \NIV{}. We find that the \NV{} $\lambda\lambda1242,\,1238$ resonance doublet in our model is not significantly affected by the applied X-ray field, while the \NIV{}~$\lambda\lambda1718,\,1721$ doublet is very sensitive. However, in the region around the \NIV{}~$\lambda\lambda1718,\,1721$  doublet, the observations are very noisy and can be explained by the model with X-rays as well as without.

    The composed model spectrum as well as the individual spectra are shown in Figs.~\ref{fig:balmer}\,--\,\ref{fig:spectral_fit2}. A summary of the stellar parameters including the abundances is given in Table~\ref{tab:stellar_parameters_summary}.
    
\subsubsection{Spectroscopic stellar masses}
    The spectroscopic masses derived from our analysis are ${M_\mathrm{spec,\,1} = 29^{+17}_{-11}\,\msun}$ for the primary and  $M_\mathrm{spec,\,2} = 22^{+13}_{-8}\,\msun$ for the secondary. The spectroscopic masses, although a factor of $1.5$ higher than the orbital masses, agree with these latter within their respective uncertainties. The large error margins on the spectroscopic masses arise mainly from the large uncertainties on surface gravity and  luminosity (e.g., Fig.~\ref{fig:balmer}). The spectroscopic mass ratio is ${q_\mathrm{spec}=0.7}$ which is smaller than the one obtained from the orbital analysis. 
    
    It appears that the spectroscopically derived masses and luminosities are shifted systematically to higher values compared to the results of the orbital analysis. We discuss this in more detail in Sect.~\ref{sec:mass_discrepancy}.
    
    Nevertheless, a mass ratio close to unity, while the two stars in a close binary have different spectral types, strongly suggests mass transfer in the past.
    From our binary evolutionary models (Sect. \ref{sec:evol_analysis}), we expect that mass transfer removed most of the hydrogen-rich envelope from the  primary. We calculated model spectra with strongly depleted surface hydrogen ($X_\mathrm{H} = 0.2$ and $0.5$). However, these models yield poorer spectral fits as the helium lines become too deep. Still, moderate hydrogen depletion and helium enrichment cannot fully be excluded.
    
\subsection{Revisiting the spectral classification}
\subsubsection{Spectral classification aided by stellar atmosphere modeling}
\label{sec:spec1}
    AzV\,476 was previously classified on the basis of the 
appearance of its optical spectrum. However, during our spectroscopic analysis, we found that most of the classification lines are blended by the companion. Hence, including the additional information from our spectroscopic models and the available UV data allows us to reconsider the spectral classification.
    
    We used the Marxist Ghost Buster (MGB) spectral classification code \citep{mai1:12,mai1:19}. MGB compares observed spectra with the criteria of \citet{wal1:00,wal1:02,wal1:14} and \citet{sot1:11,sot1:14} while allowing the presence of two components by iteratively adjusting the spectral types, light ratio, RVs, and rotational indices.
This approach poses two problems. First, the low metallicity of the SMC leads to weaker nitrogen lines compared to stars with similar type in the MW and LMC. Second, only the X-SHOOTER spectrum is covering the whole wavelength range of $\SIrange{3900}{5100}{\AA}$ which is needed for a spectral classification. In addition we consider those UVES spectra with the highest RV shifts in order to better separate the individual binary components.
    
    Therefore, we are forced to use the X-SHOOTER spectrum first and then consider the UVES spectra with the highest RV shifts that cover the range from {3300} to {4500}{\AA}.
    
    The MGB code provides the spectral type O4\,IV((f)) for the primary. The main reason for the slightly later subtype assigned to the primary is its contribution to the \HeI{} $\lambda4471$ line. The suffix ``((f))'' is assigned because of the barely detectable \NIII{}~$\lambda\lambda4634,4641$ lines (see Fig.~\ref{fig:nitro2}). Regarding the secondary, the absence of the \HeII{} $\lambda4200$ line suggests the latest O subtype, O9.5:\,Vn. The suffix ``n'' refers to the high rotational broadening of the lines of the  secondary.
    
    The MGB classification code only considers the optical range of the spectrum. If one inspects the UV spectrum, which mostly originates from the primary, and compares it to the \NV{} and \CIV{} UV wind-profile templates with those presented in \citet{wal1:00}, one might prefer a luminosity class ``III'' for the primary. The difference in the optical spectral appearance is explainable by the high L/M ratio which leads to very strong mass loss and thus to exceptionally strong wind lines in the UV. In light of these circumstances and the unusual chemical composition, we add the suffix ``p'' for this peculiar object. In the end, we therefore classify the system as O4\,IV-III((f))p~+~O9.5:\,Vn.

    Following the work of \citet{wei1:10}, an isolated star with spectral type O4\,III is expected to have a stellar mass of about $M_\mathrm{exp,\,1}=49^{+7}_{-6}\,\msun,$ which is more than twice as high as that which we derive using two different methods (Sects.~\ref{sec:orbit_analyis} and \ref{sec:spec_analysis}). For the O9.5:\,V secondary, the mass expected from the spectral type would be $M_\mathrm{exp,\,2}=16^{+7}_{-3}\,\msun,$ which agrees within uncertainties with the findings of the orbital and spectroscopic analysis.
    
\section{Stellar evolution modeling}
\label{sec:evol_analysis}

    From orbital and spectroscopic analysis, we find that the mass ratio is close to unity, although the two binary components have distinctly different luminosities, effective temperatures, and rotation rates: the secondary is a fast rotator while the rotation of the  primary is only moderate. These facts strongly indicate that the system has already undergone mass transfer.
    
\subsection{Single-star models}
\label{sec:single_stars}

    As a first approach we investigate whether the stellar parameters of the individual components can be reproduced by single-star evolutionary models. For this purpose, we employ the Bayesian statistic tool ``The BONN Stellar Astrophysics Interface'' (BONNSAI\footnote{www.astro.uni-bonn.de/stars/bonnsai}), \citep{sch1:14} in combination with the BONN-SMC tracks \citep{bro1:11}. To find a suitable model, we request the tool to match the current luminosity, effective temperature, rotation rate, and orbital mass of the two stars. Indeed, we find that the fast-rotating secondary can be partially explained by a single-star track. With respect to the primary, the BONNSAI tool cannot find any track that would explain its current low mass. Only when ignoring the mass constraint are we able to find a suitable track that reproduces the remaining stellar parameters.    In that case, the predicted ages of the primary and secondary differ by a factor of two. This finding confirms our  suggestion that the system has already undergone mass transfer. The best-fitting stellar parameters obtained with the BONNSAI tool ---applicable only for single stars--- are listed in Table~\ref{tab:stellar_parameters_summary}.
    
\subsection{Binary evolutionary models}
\label{sec:binary_stars}

    \begin{figure}[t]
        \centering
        \includegraphics[trim= 0.5cm 0.6cm 2.4cm 1cm ,clip ,width=0.50\textwidth]{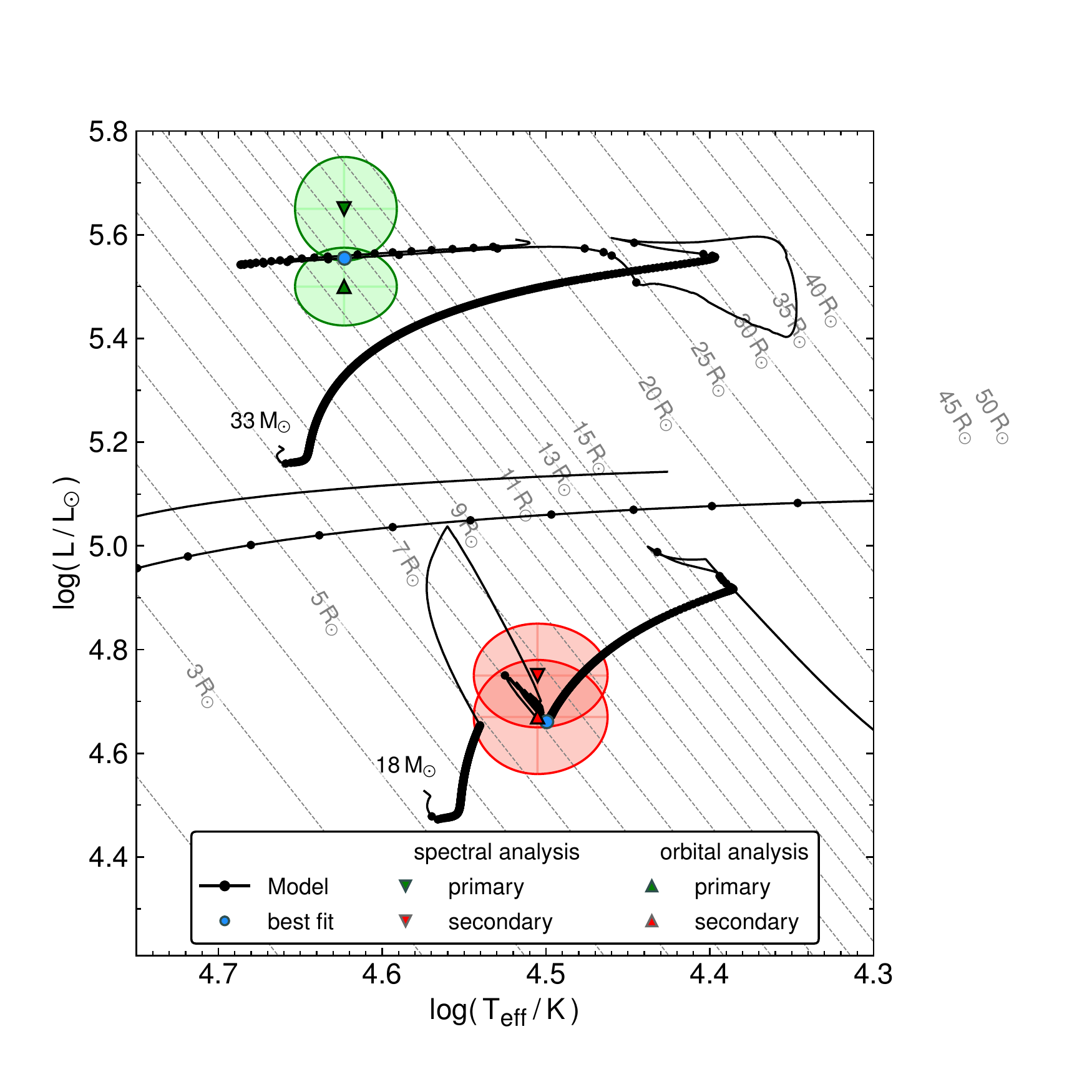}
        \caption{HRD showing the positions of each binary component according to the spectroscopic and orbital analysis, and the best fitting binary evolutionary tracks. 
        The results of the spectroscopic analysis are shown as triangles and the orbital analysis as upside-down triangles. The positions of the primary and secondary 
are marked in green and red, respectively. The shaded areas indicate the respective error-ellipses. Evolutionary tracks (solid lines) of both binary components are according to the best fitting binary model calculated with MESA (see Sect.~\ref{sec:binary_stars}). The tracks are labeled by their initial masses. The black dots on the tracks correspond to equidistant time-steps of $\SI{0.1}{Myr}$ to emphasize the most probable observable phases.  The light blue dots mark the best-fitting model.}
        \label{fig:HRD_binary}
    \end{figure} 
    
    The binary models are calculated with MESA v.\,10398. To mimic the BONN-SMC tracks we adjusted MESA in a similar way to that described by \citet{mar2:16}. We follow most of the physical assumptions from the BONN-SMC models (e.g., overshooting, thermohaline mixing, etc.) and adopt them from \citet{mar2:16} with two exceptions. In our models, we use a more efficient semi-convection with $\alpha_\mathrm{sc}=10$ and calculate the mass transfer according to the ``Kolb'' scheme \citep{kol1:90}, as it allows us to include the eccentricity enhancement mechanism meaning that the evolution of the eccentricity can be modeled properly.
    
    We want to emphasize that our goal is to check whether  binary evolutionary models can explain the stellar parameters of AzV\,476 and especially its  masses. Therefore, 
    we only explore a narrow parameter space and compute a small set of models with initial primary masses in the range of ${M_\mathrm{ini,\,1}=(\numrange{25}{38})\,\msun}$ and secondary masses ${M_\mathrm{ini,\,2}=(\numrange{10}{25})\,\msun}$, orbital periods in the range of ${P_\mathrm{\,ini}=\SIrange{8}{25}{d}}$ and initial eccentricities ${e_\mathrm{\,ini}=\SIrange{0.0}{0.4}{}}$. The initial parameters are adjusted such that at some later stage the model binary has properties similar to AzV\,476.
    We assume that the stellar components initially rotate with ${\varv_\mathrm{rot}=0.1\,\varv_\mathrm{crit}\approx\SI{65}{km\,s^{-1}}}$. These values are chosen arbitrarily but as we are dealing with a close binary, the tidal forces nevertheless lead to a tidal synchronization. Further important assumptions  are as follows: 
    
    Our models include the effect of inflation, which appears inside a stellar envelope when the local Eddington limit is reached and exceeded, leading to a convective region and a density inversion \citep{san1:15}.
    
    The stellar wind prescription is inspired by the work of \citet{bro1:11}. The winds of hot H-rich stars are described according to  the \citet{vin1:01} recipe. For stars with effective temperatures below the bi-stability jump, where mass-loss rates abruptly increase \citep[see][]{vin1:01}, we use the maximum  $\dot{M}$ from either \citet{vin1:01} or \citet{nie1:90}. WR mass-loss rates are according to \citet{She1:19} but are assumed to scale with metallicity as $\dot{M}\propto Z^{1.2}$ as recommended by \citet{hai1:17}. In the transition phase from a hot H-rich star ($X_\mathrm{H} \geq 0.7$) to the WR stage ($X_\mathrm{H} \leq0.4$), the mass-loss rate is interpolated between the prescriptions of \citet{vin1:01} and \citet{She1:19}.

    Rotational mixing is modeled as a diffusive process including the effects of dynamical and secular shear instabilities, the Goldreich-Schubert-Fricke instability, and Eddington-Sweet circulations \citep{heg1:00}. In addition to the angular momentum transport by rotation, the transport via magnetic fields from the Tayler-Spruit dynamo \citep{spr1:02} is also included.

    Convection is described according to the Ledoux criterion and the mixing length theory \citep{boe1:58} with a mixing length parameter of ${\alpha_\mathrm{mlt}=l/H_\mathrm{P}=1.5}$. For hydrogen burning cores, a steep overshooting is used such that the convective core is extended by ${0.335H_\mathrm{P}}$ \citep{bro1:11} where ${H_\mathrm{P}}$ is the pressure scale height at the boundary of the convective core. Thermohaline mixing is included with an efficiency parameter of ${\alpha_\mathrm{th}=1}$ \citep{kip1:80}, as well as semiconvection with an efficiency parameter of ${\alpha_\mathrm{sc} = 10}$ \citep{lan1:83}.
    
    Mass transfer in a binary is modeled using the ``Kolb'' scheme \citep{kol1:90}. This allows us to use the Soker eccentricity enhancement \citep{sok1:00}, which assumes phase-dependent mass loss and calculates the change in eccentricity due to the mass that is lost from the system, and the mass that is accreted by the companion. Tidal circularisation is taken into account throughout the entire evolution of the system.
    
    The remaining mass collapses into a compact object ---represented by a point mass--- when helium is depleted in the stellar core. This simplification avoids numerical problems in the latest burning stages and the unknowns that come with a supernova explosion and a possible kick.

    As the luminosities obtained from the spectroscopic and orbital analysis differ to some extent, we put additional focus on reproducing the orbital masses, which are our most reliable estimate.
	The best agreement with the empirically derived parameters is found for a system with initial masses ${M_\mathrm{ini,\,1}=33\,\msun}$ and $M_\mathrm{ini,\,2}=17.5\,\msun$, initial orbital period ${P_\mathrm{\,ini}=\SI{12.4}{d}}$, and initial eccentricity  ${e_\mathrm{\,ini}=0.14}$. The model closest to the current stellar parameters gives similar masses for the primary and secondary: ${M_\mathrm{evol,\,1}=17.8\,\msun}$ and ${M_\mathrm{evol,\,2}=18.2\,\msun}$,.    The remaining stellar parameters of our favorite models are listed in Table~\ref{tab:stellar_parameters_summary}.
    
    The Hertzsprung-Russell diagram (HRD) displaying the best fitting stellar evolutionary tracks for both binary components is illustrated in Fig.~\ref{fig:HRD_binary}. 
    The binary evolution model is able to reproduce almost all empirically derived stellar parameters including the current orbital period (${P_\mathrm{model}=\SI{9.3}{d}}$) and the eccentricity of the system (${e_\mathrm{model}=0.25}$). However, the evolutionary models over-predict the rotation rate of the secondary and the surface abundances of the primary; see Table~\ref{tab:stellar_parameters_summary}. We calculated PoWR atmosphere models with the best fitting parameters obtained with our MESA models. The synthetic spectra are shown in Fig.\,\ref{fig:spectral_fit_MESA}.
    
    A more fine-tuned model would likely solve some of these problems. For example, we did not include the initial rotation as a free parameter. This might solve the faster rotation of the primary; however, the secondary would still be expected to be rotating too fast as it spins up to criticality during mass transfer.
    According to our favorite evolutionary model, the primary in AzV\,476 is currently evolving towards the helium zero age MS (ZAMS) and will possibly spend the rest of its life as a hot helium or WR-type star.
    
\section{Discussion}
\label{sec:disc}
    
\subsection{Comparison of the orbital, spectroscopic, and evolutionary mass estimates }
\label{sec:mass_discrepancy}

    The orbital masses and the masses predicted by the binary evolutionary models agree well within their error margins and deviate only by $10\%$.  On the other hand, the spectroscopic mass and luminosity of both components appear to be higher by a factor of $1.5$ compared to the orbital solutions. 
    The question therefore arises as to whether the discrepancy between the orbital and spectroscopic solutions is significant.
    
\subsubsection{Orbital versus spectroscopic mass}

    The simplest way to resolve this discrepancy is to assume that the distance to the system is lower than the canonical SMC distance of (${d=\SI{55}{kpc}}$). As the luminosity depends on the assumed distance as $L\propto d^{-2}$, the radius (${R\propto\sqrt{L}\propto d^{-1}}$) and the spectroscopic mass  also depend on distance (${M_\mathrm{spec}\propto R^2\propto L \propto d^{-2}}$).
    The SMC galaxy is extended, and the distances to its various structural parts are not well established.  \citet{tat1:20} suggest that young structures, such as the NGC\,456 cluster, are not well represented by the intermediate age stellar populations, which are usually used for distance estimates. Instead, young clusters might be located in front of these latter. This is in agreement with the argumentation by \citet{ham1:15}, who showed that the interactions between the LMC and SMC could lead to multiple tidal and ram-pressure stripped structures, which spatially separate young star-forming regions from older stellar populations.  
    
    In order to bring the spectroscopic and orbital parameters of AzV~476 into accordance, the system needs to be shifted to a distance of $\SI{49}{kpc}$. The resulting spectroscopic masses are then ${M_\mathrm{shift,\,1}}\approx 23\,\msun$ and ${M_\mathrm{shift,\,2}\approx 18\,\msun}$. 
    However,  $d=\SI{49}{kpc}$ would imply that our target is located at the same distance as the LMC.
    
    The LMC has higher metallcity (${Z_\mathrm{LMC}=1/2\,\zsun}$) than the SMC (${Z_\mathrm{SMC}=1/7\,\zsun}$), and therefore we speculate that stars formed in the interaction regions may have higher metallicity than the SMC. To test this,   
    we increased the metallicity content of our spectroscopic models to the LMC metallicity while keeping the remaining stellar parameters unchanged.  We find that the lines in the iron forest, especially in the far-UV, are too deep compared to the observations. Therefore, it cannot be confirmed that our target has LMC metallicity, but at the same time this does not provide additional constraints on its distance.

    Another option we need to consider is the possibility of a third light contribution. AzV\,476 is located in a very crowded region. Therefore, it is possible that the observed light is contaminated by a nearby object and hence the true luminosity (and mass) is lower. We inspected the HST acquisition image of our target and find no other nearby UV-bright stars. Alternatively our target could be a multiple system rather than a binary. If there were a third component, it would need to be of similar luminosity to the secondary in order to have an impact on our analysis. However, we do not see any indications of a third component in our spectra.
    
    A third option is that the surface gravities determined from the spectral analysis are overestimated. The accuracy of the spectroscopic mass strongly depends on surface gravity, $\log\,g$, which is mainly determined from fitting the pressure-broadened line wings. Such estimates suffer from limited $S/N$ ratio in the spectra. In cases where the Balmer lines include contributions from both components, further uncertainties are induced. As explained in Sect.~\ref{sec:spectral_fitting} and shown in Fig.~\ref{fig:balmer},
    the \HeII{} lines help to improve the estimates, but still do not allow us to pin point  $\log\,g$ for each component separately. This leaves room for various combinations of $\log\,g$ able to reproduce the Balmer lines and is reflected in the large uncertainties of the spectroscopic masses.
    
\subsubsection{Orbital versus evolutionary mass}

    The evolutionary modeling yields a slightly lower mass for the primary than the spectroscopic and orbital analyses. As can be seen in Table~\ref{tab:stellar_parameters_summary}, the evolutionary mass is  $~2\,\msun$ lower than the orbital one.
    
    For a given initial mass, the main parameter defining how much mass is lost during mass transfer by the primary is the initial binary orbital period. As a rule of thumb, the longer the initial orbital period, the less mass is removed from the donor star \citep{mar2:16}. However, several effects (e.g., eccentricity) influence the response of the orbit to mass transfer. For instance, an eccentric orbit leads to phase-dependent mass transfer, which potentially increases the eccentricity and widens the orbit. Thus, the donor can expand more, and less mass is removed via Roche-lobe overflow \citep{sok1:00,vos1:15}.
    
    The amount of mass lost during mass transfer is also reduced when the radial expansion of the star is minimized such that the star can stay underneath the Roche lobe. The radial expansion is affected by the mixing processes. For instance, a star with more efficient semi-convective mixing is expected to be more compact \citep[e.g.,][]{kle1:20,gil1:21}. We computed a binary evolutionary model with a less efficient semi-convection of ${\alpha_\mathrm{sc}=1}$ and find that the post-mass-transfer (after detachment) mass of the primary is lowered by $1\,\msun$. This is only one example and there are a handful of  other processes that impact the mixing efficiency, energy transport, and the expansion of the envelope.
    
    One more process that affects the radial expansion is the mass-loss rate which is described by the adopted mass-loss recipe (see Sect.~\ref{sec:evol_analysis}).  Over- or underestimating the mass-loss rate influences stellar evolution before, during, and after mass transfer. Interestingly, we find approximate agreement  between the empirically derived mass-loss rate of the primary and the prediction of the wind recipe (see Sect.~\ref{sec:primary_mdot}).
    
    The evolutionary mass of a star after mass transfer is sensitive  to various assumptions. Therefore, a difference of $2\,\msun$ between evolutionary and orbital mass appears to be within the range of the evolutionary model uncertainties. While a more fine-tuned binary evolutionary model is beyond the scope of this paper, our modeling confirms that the orbital masses of AzV\,476 are well explained by a post-mass-transfer binary evolutionary model. Therefore,  this unique system is a useful benchmark for improving the understanding of mass transfer and mixing processes in close massive binaries.
We conclude that the orbital, spectroscopic, and evolutionary mass estimates agree within their uncertainties.

\subsection{The empirical mass-loss rate} 

    Stellar evolution depends on the mass-loss rate. Stellar evolution models rely on recipes that are only valid for a given evolutionary phase. In transition-phases one typically interpolates between corresponding mass-loss recipes (e.g., Sect.~\ref{sec:evol_analysis}). In the following, we  compare the mass-loss rates  we empirically derive from spectroscopy with other estimates as well as with the various recipes.

    In this work, the mass-loss rates are primarily derived from resonance lines in the UV and the \Halpha{} line while taking into account the morphology of the \HeII{}~$\lambda1640$ and \HeII{}~$\lambda4686$ lines.  The winds of all hot stars are clumpy \citep[e.g.,][]{ham1:08} which enhances the emission lines fed by recombination cascades \citep[e.g.,][]{ham1:98} such as the \Halpha{} line. Resonance lines in the UV are mostly formed by line scattering, a process that scales linearly with density and is thus independent from microclumping. Considering both \Halpha\ and the resonance lines allows us to constrain the clumping factor $D$ and the mass-loss rate $\dot{M}$ consistently.

\subsubsection{The primary}
\label{sec:primary_mdot}
    The spectroscopically derived mass-loss rate of the primary is ${\log(\dot{M}/(\msunpyr))=-6.1\pm0.2}$. \citet{mas1:17} derive a mass-loss rate of $\log(\dot{M}/(\msunpyr))=-5.54^{+0.12}_{-0.10}$ from the IR excess. However, the clumping effects were not included in their analysis. As the IR excess originates from free-free emission, 
    it scales with $\sqrt{D}$. Correcting the mass-loss rate with a clumping factor of $D=20$, as assumed in our spectroscopic models, yields $\log(\dot{M}/(\msunpyr))=-6.19^{+0.12}_{-0.10}$,  which is in agreement with the optical and UV analyses.
    
    In the massive star evolutionary models, the terminal wind velocity is prescribed as   $\varv_\infty/\varv_\mathrm{esc}=2.6$ and the mass-loss rates of OB stars are prescribed according to   
    the \citet{vin1:01} recipe. Using the stellar parameters obtained with spectroscopic analysis (see Table~\ref{tab:stellar_parameters_summary}) and the prescribed $\varv_\infty=\SI{1911}{km\,s^{-1}}$, the ``Vink's recipe'' yields ${\log(\dot{M}/(\msunpyr))=-6.1}$. Alternatively, applying the actual wind velocity measured from the UV spectra, $\varv_\infty=\SI{2500}{km\,s^{-1}}$, the theoretical prediction changes to ${\log(\dot{M}/(\msunpyr))=-6.2}$, which is in agreement with our empirical result within the uncertainties.
    
    However, the primary's current mass-loss rate adopted in our favored binary evolutionary model (${\log(\dot{M}/(\msunpyr))=-6.4}$) is lower. This is mainly caused by the interpolation between the mass-loss rate prescriptions for OB stars by \citet{vin1:01} and for WR stars by \citet{She1:19} which is employed when the surface hydrogen mass-fraction of the primary's evolutionary model already dropped below $X_\mathrm{H}<0.7$. Additionally, the stellar parameters (e.g., mass and luminosity) that enter the different mass-loss recipes employed by the evolutionary code are slightly different than those obtained from our spectral analysis.
    
    Based on dynamically consistent Monte Carlo wind models, \citet{vin1:21} suggest an updated scaling of metallicity for their mass-loss rates. This scaling yields higher mass-loss rates of  ${\log(\dot{M}/(\msunpyr))=-5.7}$, or, using the measured  $\varv_\infty=\SI{2500}{km\,s^{-1}}$, ${\log(\dot{M}/(\msunpyr))=-5.9}$.  Correcting in a simple way the former theoretical mass-loss rate for clumping results in  ${\log(\dot{M}/(\msunpyr))=-6.4}$, which is somewhat lower than our empirical measurement. 
    
    \begin{figure*}[htpb]
        \centering
        \includegraphics[trim= 0.8cm 1cm 1.8cm 0cm ,clip ,width=0.49\textwidth]{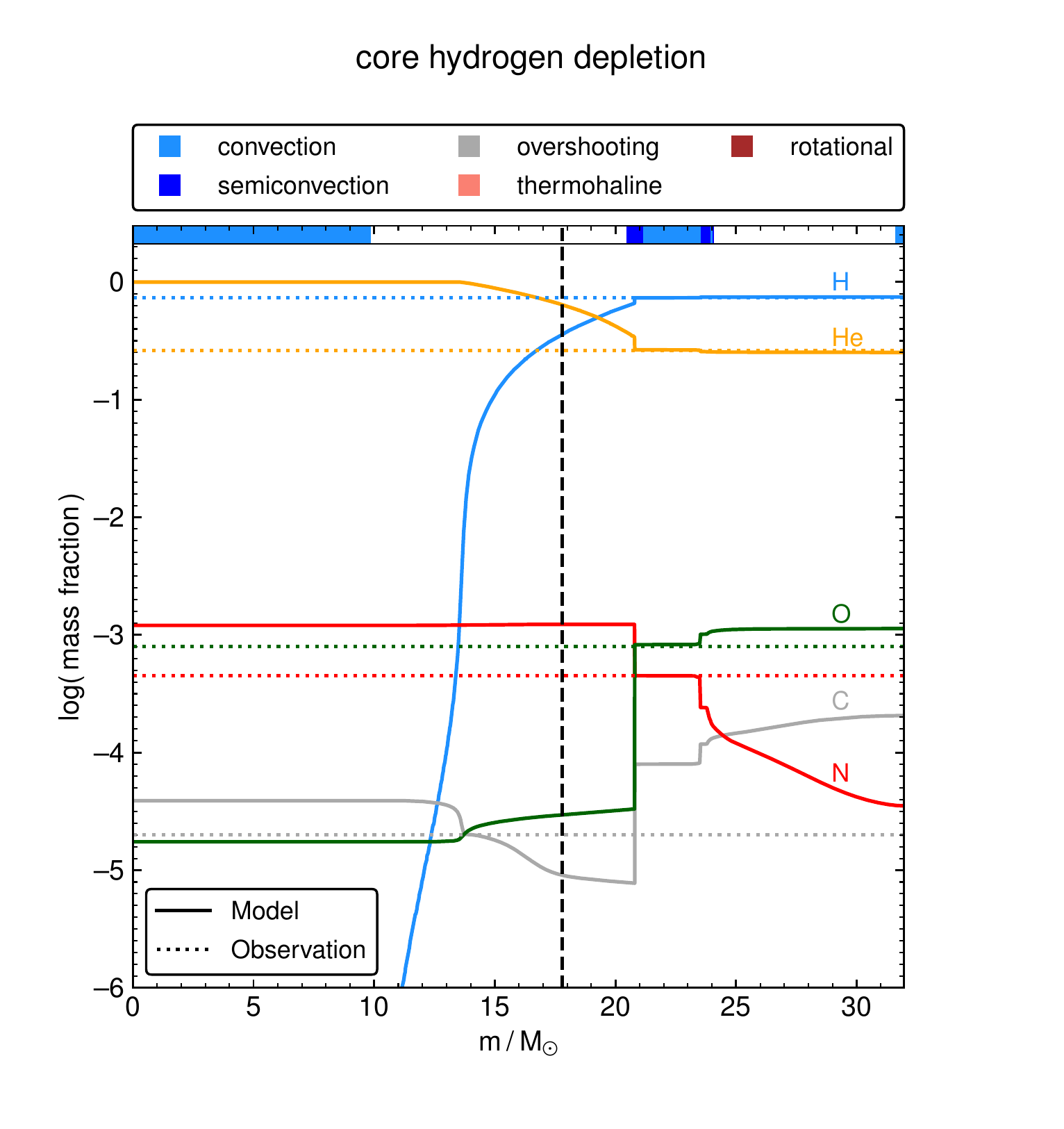}
        \includegraphics[trim= 1.8cm 1cm 0.8cm 0cm ,clip ,width=0.49\textwidth]{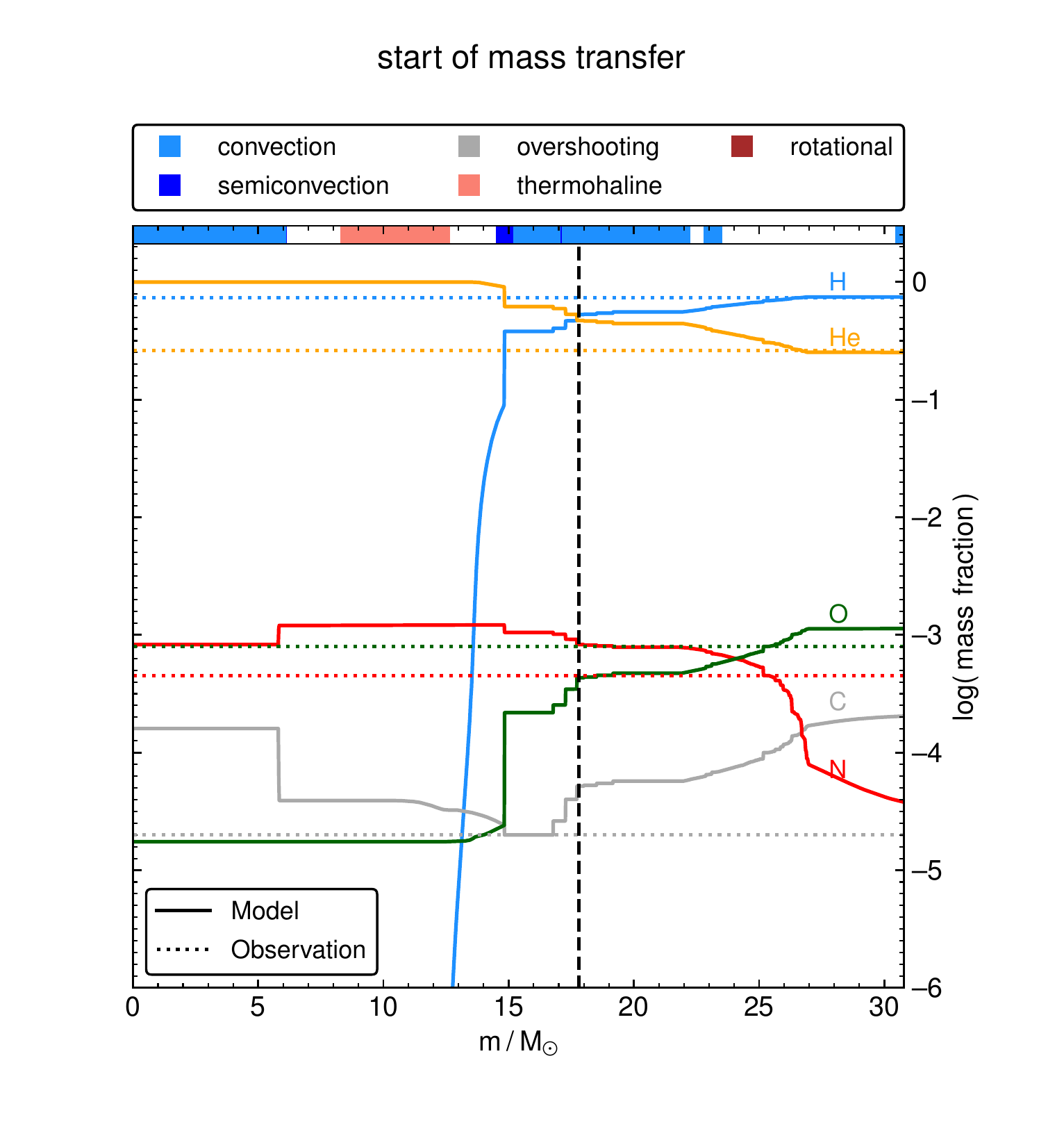}
        \caption{Abundance profiles of the primary at different evolutionary stages as predicted by the evolutionary models. The legend at the top of each plot explains the colors indicating the dominating mixing processes. The H, He, C, N, and O abundances of the model are shown as solid lines while the observed surface CNO abundances are shown as dotted lines of the same respective color. The vertical black dotted line indicates the mass of the  primary after mass transfer. \textit{Left panel:} Primary has depleted hydrogen in the core and is at the terminal age MS. \textit{Right panel:} Primary fills its Roche lobe and starts transferring mass.}
        \label{fig:abundances}
    \end{figure*}
    
    The  spectroscopic measurements of mass-loss rates of  O-dwarfs in the SMC deliver useful empiric prescriptions \citep[e.g.,][]{bou1:03,ram1:19}. According to \citet[][see their figure\,15]{ram1:19}, the mass-loss rate of the primary star in AzV~476 would be only ${\log(\dot{M}/(\msunpyr))=-7.4}$, which is much lower than obtained from our analysis. We explain this discrepancy by the advanced evolutionary status: the primary  has already a substantially reduced mass due to its close-binary evolution, leading to a high L/M ratio. This precludes the comparison with the recipes which use the mass--luminosity relations for single stars.

\subsubsection{The secondary}
    
    There are no indications of wind lines of the secondary in the observed spectra. Therefore, we can only set an upper limit on the mass-loss rate ${\log(\dot{M}/(\msunpyr))\leq-8.8\pm0.5}$. 
    The \citet{vin1:01} recipe predicts a higher value of ${\log(\dot{M}/(\msunpyr))=-8.2}$. 
    Correcting for clumping with $D=20$ yields ${\log(\dot{M}/(\msunpyr))=-8.8}$ which is in  agreement with the empirical result.
    
    According to \citet{vin1:21}, the predicted mass-loss rate of the  secondary is ${\log(\dot{M}/(\msunpyr))=-7.8}$;   correcting it for clumping  yields  ${\log(\dot{M}/(\msunpyr))=-8.4}$, which is higher than the empirical upper limit.
    The empirical relation of \citet[][their figure\,15,]{ram1:19} suggests ${\log(\dot{M}/(\msunpyr))=-8.6\pm0.2}$, which is somewhat higher than our result but consistent with its error margins.
    
    As can be seen in Table~\ref{tab:stellar_parameters_summary}, the mass-loss rate used in the binary evolution calculation is higher than predicted by the recipes as well as empirically measured. This is because
    in the MESA models, the mass-loss rate of a star rotating close to critical is enhanced compared to slowly rotating stars \citep{pax1:13}. The synthetic spectrum calculated with the parameters obtained by the binary evolutionary models displayed in Fig.\,\ref{fig:spectral_fit_MESA} shows strong P\,Cygni profiles in the spectrum of the secondary (e.g., \SiIV{}\,$\lambda\lambda1393.8, 1402.8$) that are not observed. 
    We conclude that the mass loss of the secondary is not rotationally enhanced above the observed limit.

\subsection{Comparison of the stellar parameters as predicted by MESA evolutionary tracks versus derived spectroscopically }
    
    In this work,  AzV\,476 is studied using two different approaches. (1) The empiric approach is to model the observed light curves and spectra using the PHOEBE code and the stellar atmosphere model PoWR in order to derive orbital, stellar, and wind parameters. (2) The evolutionary modeling approach is to use the MESA code to compute tracks reproducing the current stellar parameters of AzV\,476 components (Fig.~\ref{fig:HRD_binary}) simultaneously  with its orbital parameters. Below we compare the outcomes of these two approaches. 
    
    In binary evolutionary models, during the mass-transfer phase, most of the hydrogen-rich envelope is stripped off such that products of the CNO burning process are exposed at the surface. The predicted surface abundances in our favorite model correspond to an intermediate stage between the initial CNO abundances and the CNO equilibrium at metallicity ${Z_\mathrm{SMC}=1/7\,\zsun}$. This means that most of the initial C, N, and O become N, and the CO abundances are depleted.
     
    Indeed, the abundances derived spectroscopically  are intermediate  between the initial CNO abundances and the CNO equilibrium. The C and O abundances are reduced while the N abundance increased drastically. However, we find a factor of two difference between the predicted and observed N and O abundances (see Table~\ref{tab:spectroscopy_and_RV}). We note that the \emph{surface} CNO abundances predicted by the MESA models are subject to various assumptions; for example, on mixing efficiencies or the mass removal by Roche-lobe overflow. Because of the scarcity of suitable photospheric absorption lines, the CNO abundances of the primary are partially derived from emission lines. Those lines are sensitive to many effects, such as the mass-loss rate, temperature, clumping, and turbulence, introducing additional uncertainties in the abundance measurement.
    
    Furthermore, we find disagreement between the observed and predicted H mass-fraction. In evolutionary models, a significant amount of the H-rich envelope (i.e.,\ with $X_{\rm H}\,>\,0.7$) is removed during the mass-transfer phase, revealing the products of the established chemical gradient between the core and the envelope, leading to the depletion of surface hydrogen. However, from the spectroscopic analysis, we do not find such strong H depletion. As in the case of the CNO measurements, this indicates some deficiencies in the evolutionary models that are likely related to the mass-transfer phase and/or to the mixing.
    
    For a better understanding, we are showing in Fig.\,\ref{fig:abundances} abundance profiles at two different stages. The left panel in Fig.\,\ref{fig:abundances} shows the abundances at core hydrogen depletion. At this stage, the star is hydrogen shell burning, which leads to the formation of an intermediate convection zone (ICZ, \citet{lan1:87}), which ranges from the mass coordinates $\SIrange{21}{24}{\msun}$. One can see that the H, He, O, and N abundances in this region are the same as the abundances that we observe; only the predicted C abundance is higher.   However, the spectroscopically derived C abundance is the least accurate one, as the observed spectra have no photospheric C absorption lines and the \CIV{} resonance line in the UV highly depends on the mass-loss rate. At this evolutionary stage, the envelope is not well mixed, which can be seen in the step-like structure between the different mixing regions.
    
    The right panel in Fig.\,\ref{fig:abundances} shows the abundances at the onset of mass transfer. Mass transfer happens on the dynamical timescale, while the mixing and burning processes take place on longer timescales. Therefore, this plot can be used to decipher the surface composition of  the final model if a specific amount of mass is removed from the star. Compared to the left panel, the stellar model has had more time to mix the CNO elements through the envelope. The ICZ is now largely extended and the CNO abundance ratios have changed; for instance N is now the most abundant element in this region. We would like to emphasize that this highly depends on the assumed mixing efficiencies, as a less efficient semi-convection (e.g., $\alpha_\mathrm{sc}=1$) would lead to less mixing and might preserve the step-like structure seen the left panel. Semi-convection does not only affect the mixing efficiency but also impacts the change of the stellar radius \citep{kle1:20}, and therefore a star with more effective semi-convective mixing spends more time hydrogen shell burning before it fills the Roche lobe and initiates mass transfer. This additionally impacts the final chemical composition.
    
    By now it should be evident that accurate prediction of the surface CNO abundances in binary evolutionary models is a nontrivial task and is beyond the scope of this paper. However, this underlines the importance of AzV~476, which can be used to get a better understanding of the different mixing efficiencies, as it reveals the abundances formed by burning stages that would normally be hidden from the observer, making them difficult to calibrate.
        
    In order to understand how the predicted stellar parameters of the binary evolution models --- including the surface abundances --- compare to the observed spectrum, synthetic spectra of the primary and secondary are calculated using these parameters (see Table~\ref{tab:stellar_parameters_summary}).
    The resulting spectrum is shown in Fig.\,\ref{fig:spectral_fit_MESA}. When comparing the synthetic spectrum to the observed one, it is evident that the N emission lines are too strong and the helium absorption lines are too deep.
    
    The abundances are not the only difference between the evolutionary model predictions and the observations. Evolutionary model predicts much higher rotation rates than observed. In the evolutionary model, the secondary has spun up to critical rotation, which is not observed, and also the rotation rate of the  primary is somewhat slower than observed. \citet{van1:18} and \citet{sha1:20} studied the rotation rates of O stars in WR+O star binaries at different metallicities and compared them to predictions from  evolutionary models. This comparison revealed that the observed rotation rates of the O stars are lower than the predictions from the  evolutionary models. These latter authors suggest that there is a process during the fast case A mass transfer that removes the angular momentum of the star; their favored explanation is an angular momentum loss induced by a magnetic field that only develops during the mass-transfer phase.
    
    The disagreements on the chemical abundances and the rotation rates between evolutionary models and the empiric analysis suggest some unconsidered physics during the mass-transfer phase. A more fine-tuned model would likely solve some of these problems. For example, we did not include the initial rotation as a free parameter. This might solve the faster rotation of the primary; however, the secondary would still be expected to be rotating too fast as it spins up to criticality during mass transfer.
    
\subsection{AzV\,476 and its future evolution in the context of the most massive star population in the SMC}

    The binary components of AzV\,476 are among the most luminous O stars in the entire SMC. Nevertheless, our analysis reveals that the mass of the O4\,IV-III((f))p primary star is only  $M_\mathrm{orb,\,1}\approx 20\msun$.  We estimate its initial mass to be $M_\mathrm{ini,\,1}\approx 33\,\msun$ which is significantly lower than the most massive O stars in the Galaxy and the LMC \citep[e.g.,][]{ram1:17, gru1:19}. This system therefore further exacerbates the problem with the deficiency of very massive O stars in the SMC \citep{ram1:19, sch1:21}.
    
    In the future, the primary in AzV~476 will evolve bluewards on the HRD, but it will never be able to lose its entire H-rich envelope. As its luminosity is on the edge of the least luminous observed WR star of
the SMC \citep{she1:16,She1:20}, it is unclear as to whether or not it will be able to develop an optically thick wind and become a WR star. Our evolutionary model predicts that the primary will  have a final mass of $M_\mathrm{final,\,1} = 16\,\msun$ at core helium exhaustion.   The primary will probably collapse and form a BH \citep{heg1:03}. To our knowledge, no BH has yet been identified in the SMC. 
    
    After the primary collapse and the formation of a compact object, the system will consist of a MS star with a compact object. The MESA model predicts that the secondary will have  $T_\mathrm{eff}=\SI{28}{kK}$ and $\log(L/\lsun)=4.85$, at the halfway point of its remaining MS lifetime. We calculated a spectroscopic model of the secondary in this evolutionary stage and find that the secondary would be classified as a B0\,V star. Its critical rotation makes the system a likely progenitor of a high-mass X-ray binary (HMXB) with a Be donor star. After the stellar model evolves beyond the MS, it undergoes a mass-transfer phase, offering an additional opportunity to show up as a HMXB. After large fractions of the secondary's envelope are stripped off, it will remain as a helium star with some hydrogen left on the surface with a luminosity around $\log(L/\lsun)\approx4.9$. Using the recent estimates from \citet{san1:20} for massive hydrogen-free stars, the L/M-ratio will be significantly too low to yield a sufficiently strong stellar wind for it to become optically thick at this metallicity (even when accounting for additional hydrogen and larger radii). Thus, our secondary star will most likely never appear as a WR star.   With a final mass of $M_\mathrm{final,\,2}=7\,\msun$ at core helium depletion, the secondary is expected to explode as a supernova leaving a NS. With a final orbital period of about $\approx\SI{100}{d,}$ this system is not expected to merge within a Hubble time. The impact of a supernova kick is neglected in our binary models; nevertheless, in principle, it could change the final orbital period of the system drastically.

\section{Summary and conclusions}
\label{sec:conclusions}
    
    In this work we study the earliest known O-type eclipsing binary in the SMC, AzV~476.  We derived the masses of its companions using a spectroscopic analysis and by modeling its light- and RV curves. We compare the results with binary evolutionary models.
        
    We conducted a quantitative analysis of the multi-phase spectra in the optical and UV obtained by the ESO VLT and the {\em HST} and supplemented by the {\em FUSE} spectroscopy in the far-UV.  To this end we used  the non-LTE stellar atmosphere model code PoWR. The spectra of both components were disentangled, allowing us to determine their masses, stellar, and wind parameters.   Independently, we used the PHOEBE code to  derive the stellar and the orbital parameters from the light curve in the I-band as well as RV curves. We employed the MESA code to compute detailed binary evolutionary models that are able to reproduce the observed stellar parameters in significant detail (see Table~\ref{tab:stellar_parameters_summary} and Fig.~\ref{fig:HRD_binary}).
        
    \medskip\noindent
    Our conclusions can be summarized as follows: 
    {\vspace{-1ex}
    \renewcommand\labelenumi{(\theenumi)}
    \begin{enumerate}[wide,itemsep=0.5ex, labelindent=0ex]
        
    \item
    The  eclipsing binary AzV\,476 harbors one of the most luminous O stars in the SMC. It consists of an O4\,IV-III((f))p type primary and an O9.5:\,Vn type secondary. Single stars with similar spectral types typically have masses of ${M_\mathrm{exp,\,1}}=49\msun$ and ${M_\mathrm{exp,\,2}}=16\msun$.
        
    \item
    By analyzing the light curve and the RV curve 
    consistently, we derive orbital masses of ${M_\mathrm{orb,\,1}=20\msun}$ and ${M_\mathrm{orb,\,2}=18\msun}$ for the binary components. We find that the primary has less than half of the mass that is expected from its spectral type. The modest mass of one of the most luminous stars in the SMC highlights the conspicuous deficiency of very massive O-type stars in this metal-poor galaxy.
    
    \item
    The spectroscopic masses agree within the uncertainties with the more reliable orbital masses, and they confirm a mass ratio of close to unity. The observed spectra reveal that the surface N abundance of the primary is enhanced while C and O abundances are reduced. These two aspects suggest a prior mass-transfer phase.
    
    \item 
    The spectroscopic analysis uncovers that the mass-loss rate of the primary with ${\log(\dot{M}/(\msunpyr))=-6.1}$ is ten times stronger compared to recent empirical prescriptions for single O stars in the SMC. Likely, this is due to its high $L/M$ ratio.  Our empirical mass-loss rate is a factor of two higher compared to the one used in  the evolutionary models. While for our primary the result as such is in   line with standard theoretical predictions, this underlines that the    present treatment of wind mass loss in stellar evolution models needs to be improved to properly account for the products of binary evolution.
    
    \item 
    The current moderate mass of the primary can only be explained by binary evolutionary models. According to our favorite binary model, the initial mass of the primary was ${M_\mathrm{ini,\,1}=33\,\msun}$, while the secondary formed with ${M_\mathrm{ini,\,2}= 18\,\msun}$. The system is $\sim\SI{6}{Myr}$ old.  
    
    \item
    The binary evolutionary model confirms that this binary has undergone case B mass transfer and is now in a detached phase. The binary evolutionary model reveals that the primary must be core helium burning and that the observed CNO abundances correspond to those of the hydrogen shell burning layers.

    \item
    According to our binary evolution model, the primary will become a helium star (or maybe a WR star) with a portion of the hydrogen remaining in the envelope ($X_\mathrm{H}<50\%$) and finally collapse to a BH or NS. If the binary system stays bound after core-collapse, it might  show up as a HMXB with a rapidly rotating Be-type donor. After core hydrogen depletion, the secondary is expected to expand and transfer mass onto the compact object, stripping off most of its hydrogen-rich envelope. The secondary will spend most of its late evolutionary phases in the blue part of the HRD as a helium star with a small fraction of hydrogen left in the envelope. However, due to its low $L/M$ ratio, it will never be able to display a WR-like spectrum. Finally, the secondary will collapse, and there is a small chance it will form a binary of compact objects that could potentially merge within a Hubble time. 
    
    The three different methods that we used to derive the stellar masses of the binary components show that the different mass estimates agree within their respective uncertainties (e.g., Table~\ref{tab:stellar_parameters_summary} and Fig.~\ref{fig:HRD_binary}). Nonetheless, the different methods yield somewhat different results. We point out several difficulties that come along with the different methods  applied,  for instance when measuring surface gravity (see Fig.~\ref{fig:balmer}).  Finally, we conclude that  the earliest type eclipsing binary in the SMC, AzV\,476,      provides a unique laboratory for studying massive binaries at low metallicity. 
    \end{enumerate}}
    
    \begin{acknowledgements}
    We dedicate this publication to the late Dr. Rodolfo H. Barba, whose passing is a tragic loss to the astrophysics community. As the original referee of this paper, he provided an outstandingly detailed and substantial report, which was extremely helpful and led to significant improvements of this work. Dr. Barba also provided us with the extracted TESS light curve of AzV\,476 which resulted in the update in the orbital period.
    The authors thank Dr.\,Alex Fullerton for useful discussion and advise related to the 
    FUSE spectroscopy and Soetkin Janssens for very helpful advises regarding the PHOEBE code. We gratefully thank the staff of the FUSE, HST, ESO telescopes for making the useful spectroscopic data publicly available. The results presented in this paper are based on observations obtained with the NASA/ESA Hubble Space  Telescope, retrieved from the Mikulski Archive for Space Telescopes (MAST) at the 
    Space Telescope Science Institute (STScI). STScI is operated by the Association
    of Universities for Research in Astronomy, Inc. under NASA contract NAS 5-26555. Furthermore, its conclusions are based on observations collected at the European Southern Observatory (ESO) under the program 098.A-0049. The authors thank  Andrea Mehner for preparing the OBs of the XSHOOTU project and the people on the management committee of XSHOOTU.  This work has made use of data from the European Space Agency (ESA) mission {\it Gaia} (\url{https://www.cosmos.esa.int/gaia}), processed by the {\it Gaia} Data Processing and Analysis Consortium (DPAC, \url{https://www.cosmos.esa.int/web/gaia/dpac/consortium}). Funding for the DPAC has been provided by national institutions, in particular the institutions participating in the {\it Gaia} Multilateral Agreement. DP acknowledges financial support by the Deutsches Zentrum f\"ur Luft und Raumfahrt (DLR) grant FKZ 50 OR 2005.  AACS is Öpik Research Fellow at Armagh Observatory \& Planetarium. AACS and VR
    acknowledge support by the Deutsche Forschungsgemeinschaft (DFG -
    German Research Foundation) in the form of an Emmy Noether Research
    Group (grant number SA4064/1-1, PI Sander). JMA acknowledges support from the Spanish Government Ministerio de Ciencia through grant PGC2018-095\,049-B-C22.
    \end{acknowledgements}

\bibliographystyle{aa}                                                         
\bibliography{astro}           

\begin{appendix}
    
\section{RV determination}
\label{app:RV}
    
    To determine the RV shift of a star during a specific phase we use a MCMC approach. In our code we use the synthetic spectra obtained from the spectral analysis and shift them until the shifted synthetic spectrum best matches the observed spectrum. In our procedure, we use the method described in the \verb|emcee| documentation\footnote{https://emcee.readthedocs.io\label{emcee_footnote}} with small adjustments such that it fulfills our needs.
    
    The probability function, a measure of how well the synthetic spectrum fits the observed one, consists of the combination of a prior function (i.e., a function that limits the parameter space to reasonable numbers) and a likelihood function. 
      Uniform distributions are used as  prior functions to avoid biases towards specific RV shifts,
    
    \begin{equation}
        \log p_\mathrm{prior}(\varv_1,\varv_2,f)= 
        \begin{cases}
              &\text{ if } \SI{-50}{km\,s^{-1}}<\varv_1<\SI{600}{km\,s^{-1}} \\ 
            \,\,\,\,0 &\text{ and } \SI{-600}{km\,s^{-1}}<\varv_2<\SI{50}{km\,s^{-1}} \\
              &\text{ and } -\infty < \log f < 1.0\\
            \rule{0cm}{0.7cm}
            -\infty  & \text{otherwise.}
        \end{cases}
    \end{equation}
    
    For the likelihood function we use the least square function and assume that the variance is underestimated by some fractional amount $f$,
    
    \begin{equation}
         \log p_\mathrm{likelihood}=-\dfrac{1}{2}\sum_n \dfrac{(y_n-y_{\mathrm{model,\,}n})^2}{s_n^2}+\log 2\pi s_n^2,
    \end{equation} where
    
    \begin{equation}
         s_n^2 = \sigma_n^2 + f(y_{\mathrm{model,\,}n})^2.
    \end{equation}
    
    Using the above-mentioned functions, the probability function can be expressed as
    
    \begin{equation}
        \log P = \log p_\mathrm{prior} + \log p_\mathrm{likelihood}.
    \end{equation}
    
    As recommended in the \verb|emcee| documentation$^{\ref{emcee_footnote}}$ the starting values are initialized as tiny Gaussian balls around ${\varv_1=\SI{300}{km\,s^{-1}}}$, ${\varv_2=\SI{-300}{km\,s^{-1}}}$ and ${\log f = 0.0}$. To ensure a good converged posterior distribution, 32 walkers (i.e., individual Makrov chains) are used and each one is iterated for 5000 steps. We cut off the first 1000 steps to ensure that the chains are ``burned-in'' such that the remaining distribution resembles the posterior distribution. The uncertainties are based on the 16th, 50th, and 84th percentiles of the samples, corresponding to approximately $1\sigma$ deviation. To make these values readable for the PHOEBE code, which requires symmetric error margins, only the larger uncertainty is used.

    For each of the components, we take the mean value of the RVs obtained from several absorption lines.
    The primary is causing most of the \HeII{} lines seen in the composite spectrum. Only the \HeII{} $\lambda4025$ line shows contributions from both components. The secondary contributes the majority of the observed \HeI{} absorption lines, but in many of them also the primary can be seen. We use the \HeII{} $\lambda4025$ line to fit both components and to obtain their RVs simultaneously, such that this line can be used as a cross-check to see if these fits are in agreement with the other lines that are purely from the primary, like \HeII{} $\lambda4200$, or the lines that are associated with the secondary. This gives us confidence that the RVs of the secondary obtained with this method are trustworthy. The obtained RVs of the primary and secondary can be found in Table~\ref{tab:RV1} and \ref{tab:RV2}.
    
\begin{figure}[t]
        \centering
        \includegraphics[trim= 0.0cm 0cm -0.4cm 0cm ,width=0.5\textwidth]{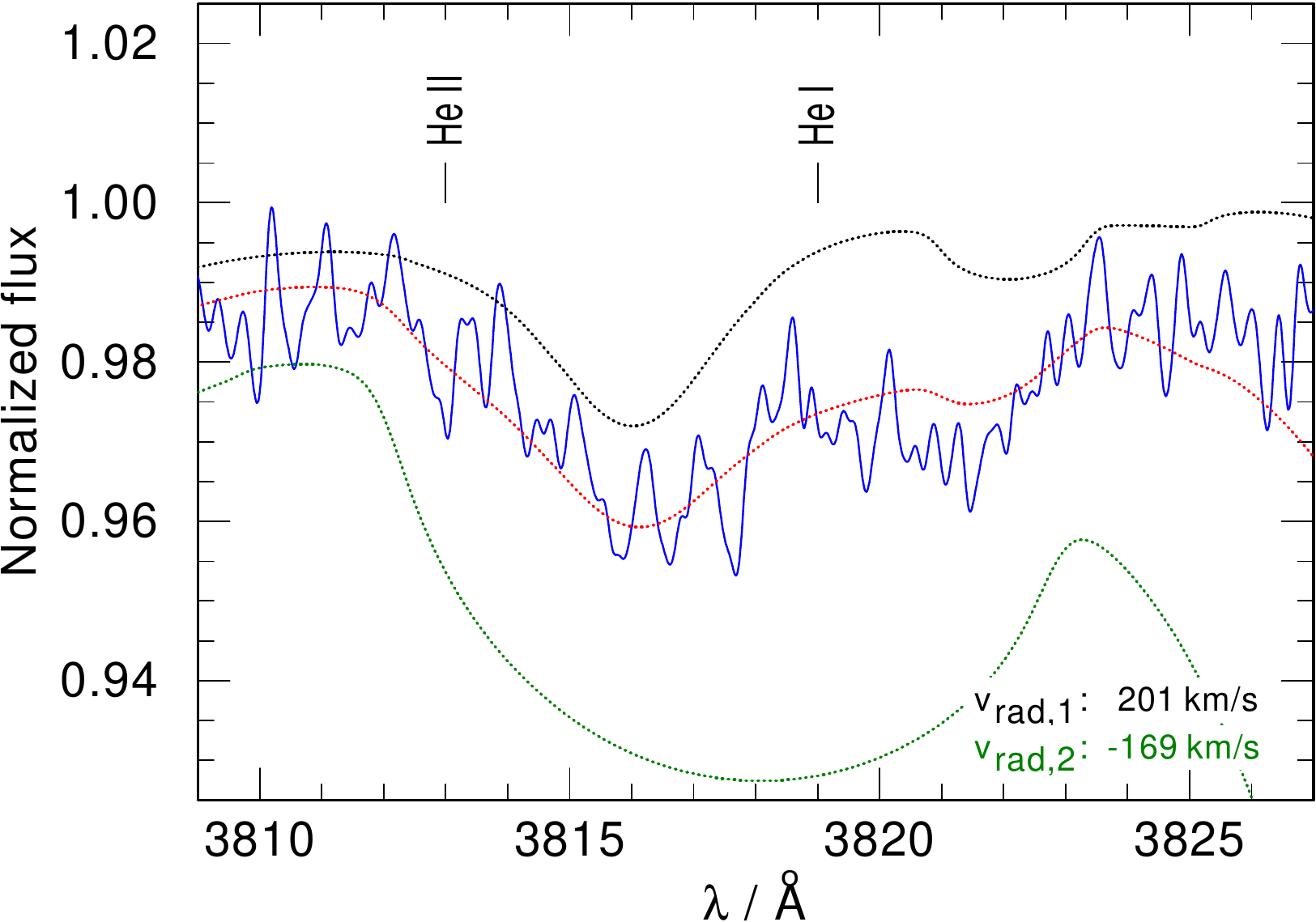}
        \caption{Radial-velocity fit of the synthetic spectrum to the observed \HeII{}\,$\lambda3813$ and \HeI{}\,$\lambda3819$ absorption lines of the UVES spectrum with ID 10 (Table~\ref{tab:spectroscopy_and_RV}).
        The observed spectrum is shown as a solid blue line, our best fitting model as a dotted red line, the unweighted synthetic spectrum of the  primary as a dashed black line, and the unweighted spectrum of the  secondary as a dotted green line. The observed spectrum is convolved with a Gaussian with a $\mathrm{FWHM}=\SI{0.2}{\AA}$ and is corrected for the velocity of the SMC and the barycentric motion. The line identification marks correspond to the wavelengths of the absorption lines in the rest frame.}
        \label{fig:heIgood}
    \end{figure}

    This fitting procedure turns out to work well even for line complexes with contributions from both stars. Figure~\ref{fig:heIgood} shows one of our fits of the \HeI{}\,$\lambda3819$ region. The \HeI{}\,$\lambda3819$ absorption line is present in both spectra. The absorption line of the  primary is shifted redwards and the rotationally broadened absorption line of the  secondary is shifted bluewards. This complex is blended by the blueshifted \HeII{}\,$\lambda3813$ absorption line of the primary.
    
    This method requires sufficiently high $\mathrm{S/N}$ observations.  The MCMC code struggles with some of the \HeI{} lines that are associated with the secondary. In most cases, the problems emerge when the depth of the absorption lines is of the same order as the noise (i.e., the UVES spectra have a $\mathrm{S/N}\sim25$). Moreover, some of the UVES spectra show a wavy pattern from the echelle orders. In some cases, these waves overlap with absorption lines, making an accurate RV determination almost impossible. All fits are inspected by eye and the observations that show the aforementioned issues are discarded from our analysis.
    
    \begin{table*}[t]
        \centering
        \caption{ RVs of the primary determined from different spectral lines.}
        \begin{tabular}{c|ccccc|c}\hline \hline \rule{0cm}{2.8ex}%
                spectral ID                   & \HeI{}\,$\lambda3819\,^{(a)}$&  \HeI{}\,$\lambda3935$& \HeII{}\,$\lambda4025\,^{(b)}$& \NIV{}\,$\lambda4058$& \HeII{}\,$\lambda4200$& mean value$^{(c)}$ \\
                                              & [$\si{km\,s^{-1}}$]          &  [$\si{km\,s^{-1}}$]  & [$\si{km\,s^{-1}}$]           & [$\si{km\,s^{-1}}$]  & [$\si{km\,s^{-1}}$]   & [$\si{km\,s^{-1}}$]\\
                \hline \rule{0cm}{3.4ex}%
                \rule{0cm}{2.8ex} 5  & $114\pm63$           &  $117\pm22$           &  $120\pm8$            &  $103\pm7$           &  $105\pm4$            &  $112\pm14$ \\
                \rule{0cm}{2.8ex} 6  & ---                  &  $229\pm26$           &  $236\pm8$            &  $217\pm21$          &  $210\pm4$            &  $223\pm9$ \\
                \rule{0cm}{2.8ex} 7  & $48\pm63$            &  $36\pm47$            &  $53\pm8$             &  $46\pm9$            &  $55\pm8$             &  $47\pm16$ \\
                \rule{0cm}{2.8ex} 8  & $140\pm13$           &  $115\pm34$           &  $134\pm11$           &  $117\pm18$          &  $122\pm4$            &  $126\pm8$ \\
                \rule{0cm}{2.8ex} 9  & $145\pm59$           &  ---                  &  $153\pm10$           &  $134\pm22$          &  $131\pm5$            &  $141\pm16$ \\
                \rule{0cm}{2.8ex} 10 & $201\pm53$           &  $215\pm25$           &  $209\pm11$           &  $195\pm18$          &  $183\pm8$            &  $201\pm13$ \\
                \rule{0cm}{2.8ex} 11 & $223\pm32$           &  $212\pm26$           &  $214\pm9$            &  $192\pm13$          &  $196\pm4$            &  $207\pm9$ \\
                \rule{0cm}{2.8ex} 12 & $162\pm18$           &  $167\pm32$           &  $131\pm14$           &  $137\pm45$          &  $144\pm8$            &  $148\pm12$ \\
                        \rule{0cm}{-1.2ex}   &                      &                       &                       &                      &                       &                          \\
                        \hline
        \end{tabular}
        \rule{0cm}{2.8ex}%
        \begin{minipage}{0.95\linewidth}
            \ignorespaces
             $^{(a)}$ The \HeI{}\,$\lambda3819$ is present in the primary and secondary spectrum. We note that the observed \HeI{}\,$\lambda3819$  line also contains the \HeII{}\,$\lambda3813$ line from the primary. $^{(b)}$ The \HeII{}\,$\lambda4025$ line has a strong contribution from the secondary. $^{(c)}$ The mean value is calculated as $\sum_{i=1}^{i=n} \mathrm{RV}_i/n$ where $\mathrm{RV}_i$ is the RV of line $i$  and $n$ is the total number of lines used. The errors of the mean values are calculated via Gaussian error propagation.%
        \end{minipage}
        \label{tab:RV1}
    \end{table*}
    
    \begin{table*}[th]
        \centering
        \caption{ RVs of the secondary determined from different spectral lines.}
        \begin{tabular}{c|ccccc|c}\hline \hline \rule{0cm}{2.8ex}%
                spectral ID                   & \HeI{}\,$\lambda3819\,^{(a)}$& \HeII{}\,$\lambda4025\,^{(b)}$& \HeI{}\,$\lambda4143$&  \HeI{}\,$\lambda4387$& \HeI{}\,$\lambda4471$&  mean value$^{(c)}$ \\
                                              & [$\si{km\,s^{-1}}$]          &  [$\si{km\,s^{-1}}$]  & [$\si{km\,s^{-1}}$]           & [$\si{km\,s^{-1}}$]  & [$\si{km\,s^{-1}}$]   & [$\si{km\,s^{-1}}$]\\
                \hline \rule{0cm}{3.4ex}%
                \rule{0cm}{2.8ex} 5  & $-88\pm29$           &  $-122\pm31$          &  ---                  &  $-117\pm20$         &  $-113\pm18$          &  $-110\pm13$ \\
                \rule{0cm}{2.8ex} 6  & ---                  &  $-164\pm22$          &  ---                  &  $-198\pm52$         &  $-152\pm24$          &  $-171\pm21$ \\
                \rule{0cm}{2.8ex} 7  & $-41\pm83$           &  $-33\pm23$           &  ---                  &  $-32\pm35$          &  $-35\pm31$           &  $-35\pm25$ \\
                \rule{0cm}{2.8ex} 8  & $-137\pm89$          &  $-79\pm41$           &  $-110\pm55$          &  ---                 &  $-91\pm24$           &  $-104\pm29$ \\
                \rule{0cm}{2.8ex} 9  & $-108\pm112$         &  ---                  &  $-95\pm88$           &  $-81\pm37$          &  $-97\pm24$           &  $-95\pm37$ \\
                \rule{0cm}{2.8ex} 10 & $-169\pm52$          &  $-199\pm43$          &  ---                  &  ---                 &  ---                  &  $-184\pm34$ \\
                \rule{0cm}{2.8ex} 11 & $-194\pm34$          &  $-151\pm26$          &  $-213\pm121$         &  ---                 &  ---                  &  $-186\pm43$ \\
                \rule{0cm}{2.8ex} 12 & $-149\pm42$          &  $-125\pm79$          &  $-146\pm68$          &  $-130\pm66$         &  $132\pm42$       &  $-132\pm27$ \\
                        \rule{0cm}{-1.2ex}   &                      &                       &                       &                      &                       &  \\
                        \hline
        \end{tabular}
        \rule{0cm}{2.8ex}%
        \begin{minipage}{0.95\linewidth}
            \ignorespaces
            Same footnotes as in Table~\ref{tab:RV1}.
        \end{minipage}
        \label{tab:RV2}
    \end{table*}
    \vfill
    \renewcommand\dbltopfraction{.5}
    \clearpage
    
\section{Additional figures}

\subsection{Additional optical \NIII{} and \NV{} lines}
\begin{figure}[htb!]
        \centering
        \includegraphics[height=0.5\textwidth,angle = -90]{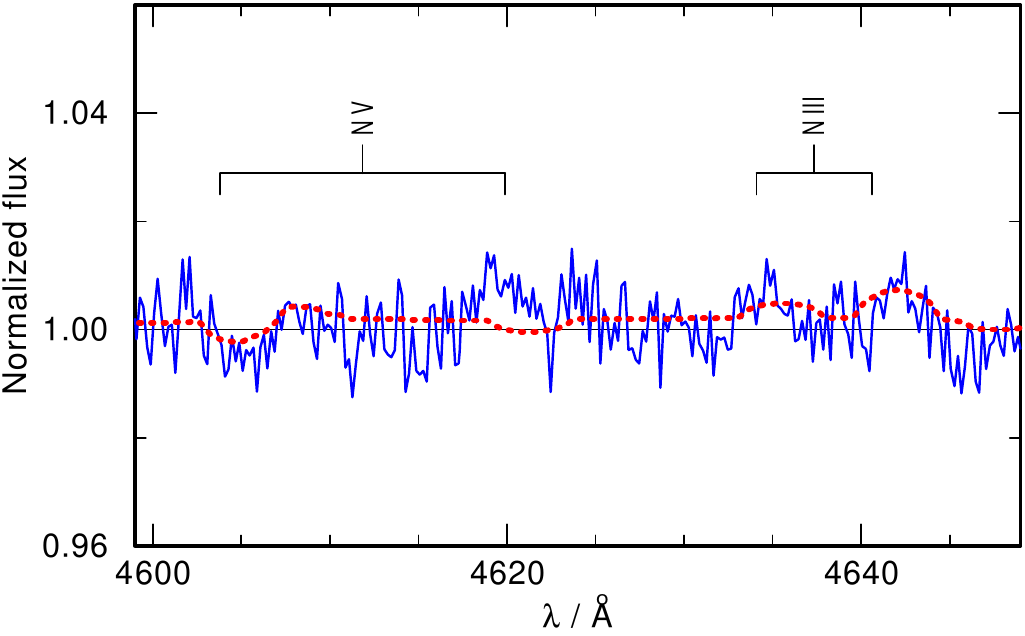}
        \caption{Same as Fig.~\ref{fig:nitro}, but now zoomed on optical \NIII{} and \NV{} lines.}
        \label{fig:nitro2}
    \end{figure}

    Figure\,\ref{fig:nitro2} shows \NIII{} and \NV{} lines in our spectrum with the highest $\mathrm{S/N}$, the X-SHOOTER spectrum (ID 5 in Table~\ref{tab:spectroscopy_and_RV}, ${\mathrm{S/N}\sim100}$). These lines are very weak and therefore barely visible in the spectrum. \NV{}\,$\lambda4604$ is marginally in absorption, while \NIII{}\,$\lambda\lambda4634,4641$ appears in slight emission. Although nearly hidden in the noise, these weak features are reproduced by our final model (red-dotted).

\subsection{Optical \OIV{} and \NIV{} lines}
\label{app:OIV}

\begin{figure}[htb!]
        \centering
        \includegraphics[height=0.5\textwidth,angle = -90]{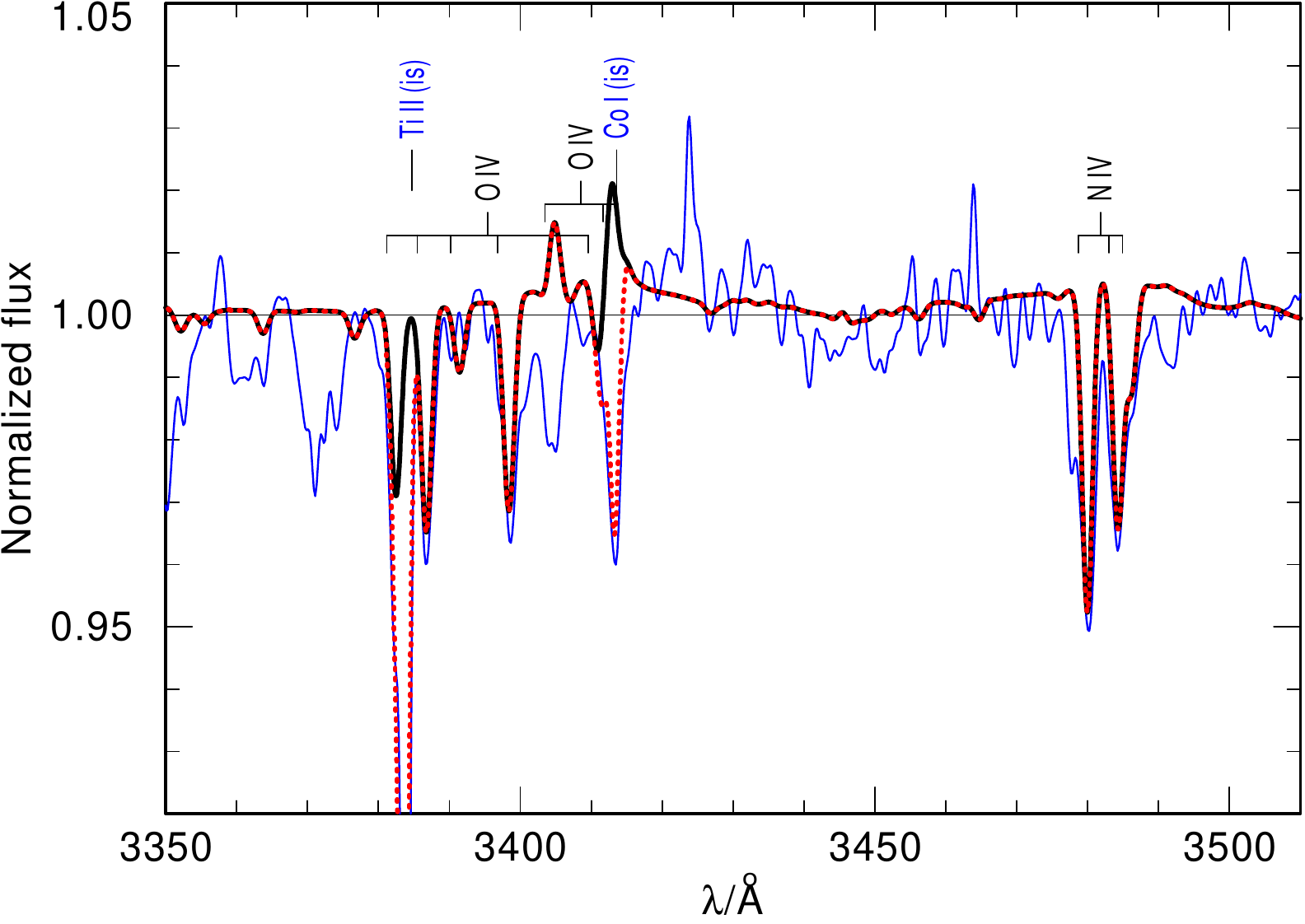}
        \caption{Observed  (X-SHOOTER spectrum; ID 5 in Table~\ref{tab:spectroscopy_and_RV}) and synthetic spectra of the region around $\SI{3400}{\AA}$ showing the \OIV{} and \NIV{} lines. The observed spectrum (blue) shows clear indications of ISM absorption lines. The black line is the synthetic spectrum without a contribution of ISM lines. The spectrum with the modeled ISM lines is shown as red dotted line. The interstellar absorption arise in the Galactic foreground and in the SMC.}
        \label{fig:oiv}
    \end{figure}

    \citet{wal1:04} suggest to use the optical \OIV{} multiplets around $\SI{3400}{\AA}$ and the \NIV{} absorption lines at wavelengths $\SI{3463}{\AA}$, $\SI{3478}{\AA}$, $\SI{3483}{\AA}$ and $\SI{3485}{\AA}$ to derive the N and O abundances in hot massive stars. Luckily, the X-SHOOTER spectrum extends to such short wavelengths. Figure~\ref{fig:oiv} shows that the \NIV{} lines of our final model are in agreement with the observation. However, for the \OIV{} multiplet at wavelengths $\SI{3403}{\AA}$ and $\SI{3414}{\AA}$ the synthetic spectrum is in emission and not in absorption as observed. For the \OIV{} $\lambda\SI{3414}{\AA}$ line,  there is an ISM contribution from \ion{Co}{I} that explains the observed absorption. Regarding the \OIV{} $\SI{3403}{\AA}$ line, we are not aware of any blending lines. However, this particular line is formed close to the onset of the wind and is sensitive to different details within the wind (e.g., temperature stratification, wind velocity law). As the other \OIV{} multiplet $\lambda\lambda \SIrange{3381}{3389}{\AA}$ can be well reproduced, we trust our determined O abundance, which is supported by other absorption lines in the UV.

\subsection{Additional wind lines}

\begin{figure}[htb!]
        \centering
        \includegraphics[height=0.5\textwidth,angle=-90]{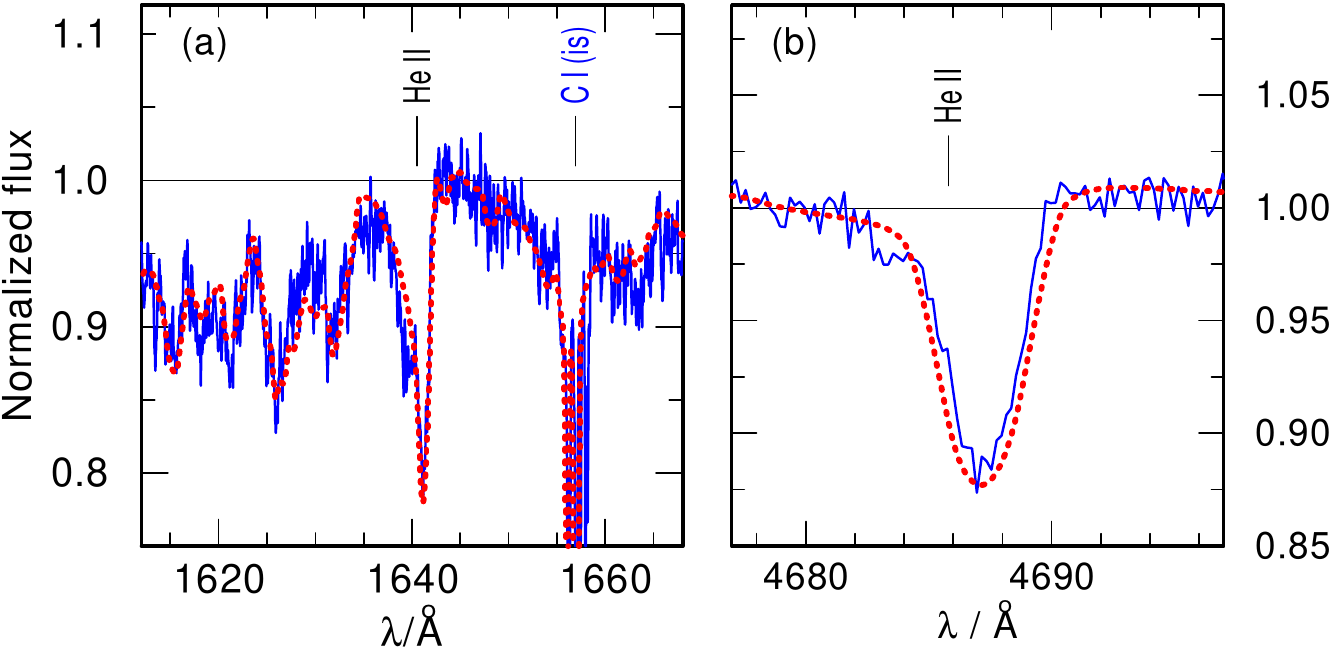}
        \caption{Observed (blue) and synthetic (red) \HeII{} lines that are sensitive to the wind parameters. The sharp interstellar absorptions arise in the Galactic foreground and in the SMC, and are also modeled with their respective RV shifts. (a) \HeII{}~$\lambda1640$ line in the COS spectrum (ID 2 in Table~\ref{tab:spectroscopy_and_RV}). (b) \HeII{}~$\lambda4868$ line in the X-SHOOTER spectrum (ID 5 in Table~\ref{tab:spectroscopy_and_RV}). }
        \label{fig:wind}
    \end{figure}

    As main diagnostic wind lines, the  \CIV{} resonance doublet in the UV and the \Halpha{} line in the optical are used. However,  the \HeII{} $\lambda1640$ in the UV and the \HeII{} $\lambda4868$ in the optical are also sensitive to the stellar wind and are taken into account when adjusting the stellar parameters of our synthetic model. These lines are shown in Fig.~\ref{fig:wind}.
    
    \clearpage
    
\onecolumn
\subsection{Synthetic spectrum calculated with the stellar parameters from the MESA model}
    \begin{figure*}[h]
        \centering
        \includegraphics[width=0.95\textwidth]{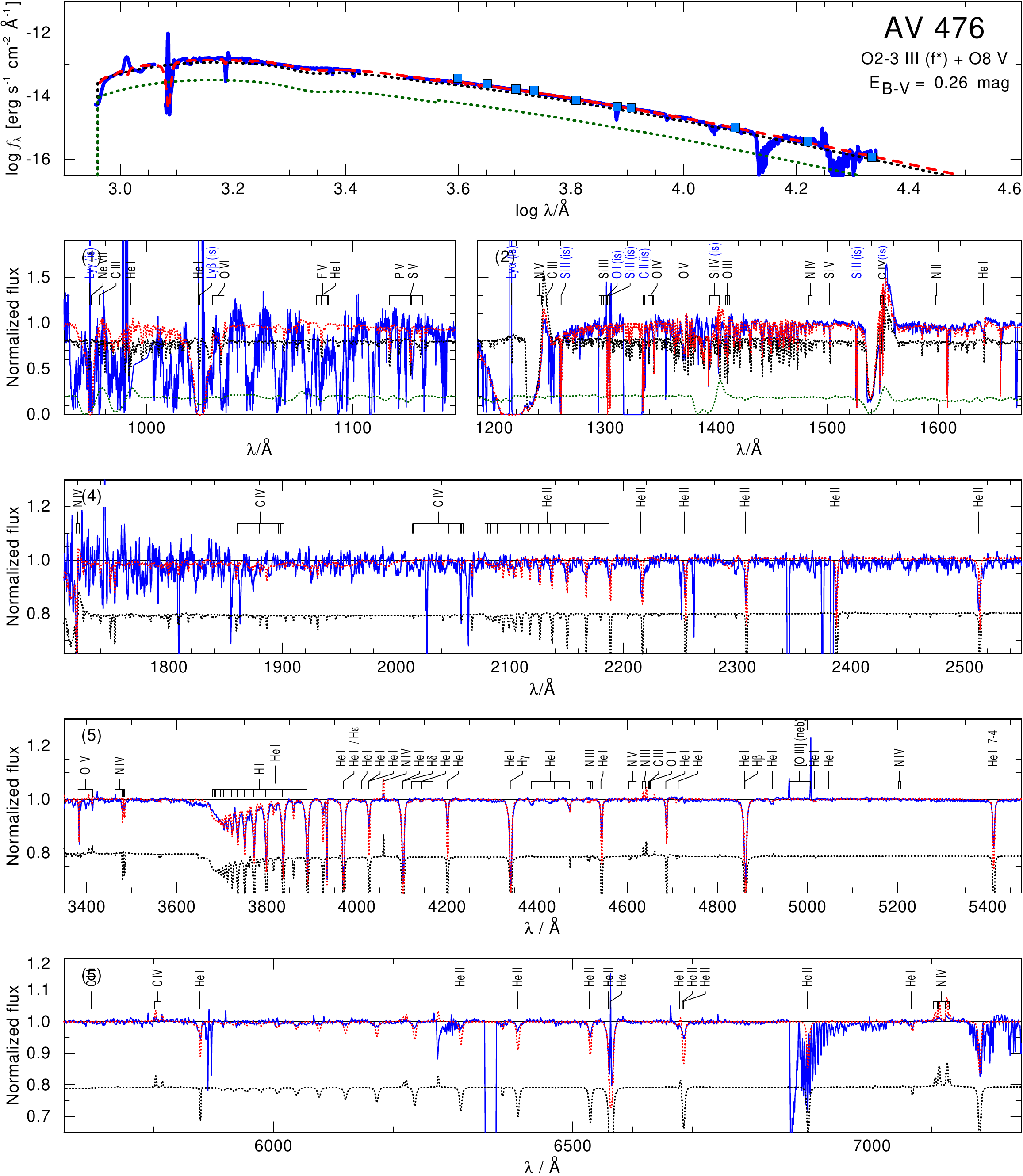}
        \caption{Spectral model calculated with the stellar parameters of the most favored binary evolutionary model. The plots and line styles are the same as in Figs.~\ref{fig:spectral_fit} and \ref{fig:spectral_fit2}. In this plot we use a distance of $d=\SI{49}{kpc}$ to compensate the used lower luminosities of the binary components.  This plot demonstrates that the predicted surface abundances of the binary evolutionary models are unable to reproduce the observed spectrum. The nitrogen abundance in this model is too strong as it overestimates all optical \NIV{} lines and also predicts unseen \NV{ } absorption. The reduced hydrogen abundance leads to an increased helium abundance. This enrichment would produce too deep \HeII{} lines in the spectrum of the  primary.}
        \label{fig:spectral_fit_MESA}
    \end{figure*}
    
\end{appendix}

\end{document}